\documentclass[journal]{IEEEtran}

\usepackage[printonlyused,withpage]{acronym}
\usepackage{cite}
\ifCLASSINFOpdf
   \usepackage[pdftex]{graphicx}
  \graphicspath{{../pdf/}{../jpeg/}}
   \DeclareGraphicsExtensions{.pdf,.jpeg,.png}
\else
   \usepackage[dvips]{graphicx}
  \graphicspath{{../eps/}}
  \DeclareGraphicsExtensions{.eps}
\fi

\usepackage{amsmath,amssymb,amsfonts}
\usepackage{algorithmic}
\usepackage{textcomp}
\usepackage{xcolor}
\usepackage{hyperref}
\usepackage{subcaption}
\usepackage{array}
\usepackage{soul}
\acrodef{ECU}[ECU]{Electronic Control Unit}
\acrodefplural{ECUs}{Electronic Control Units}
\acrodef{UNECE}[UNECE]{UN Economic Commission for Europe}
\acrodef{WP.29}[WP.29]{World Forum for Harmonization of Vehicle Regulations}
\acrodef{CAN}[CAN]{Controller Area Network}
\acrodef{CAN-MM}[CAN-MM]{CAN Multiplexed MAC}
\acrodef{CAN-FD}[CAN FD]{Controller Area Network Flexible Data-Rate}
\acrodef{CAN-XL}[CAN XL]{Controller Area Network Extra Long}
\acrodef{MAC}[MAC]{Message Authentication Code}
\acrodefplural{MACs}{Message Authentication Codes}
\acrodef{ECM}[ECM]{Engine Control Module}
\acrodef{TCM}[TCM]{Transmission Control Module}
\acrodef{ACC}[ACC]{Adaptive Cruise Control}
\acrodef{ESC}[ESC]{Electronic Stability Control}
\acrodef{ABS}[ABS]{Anti-Lock Brake System Module}
\acrodef{BCM}[BCM]{Body Control Module}
\acrodef{DEFC}[DEFC]{Diesel Exhaust Fluid Controller}
\acrodef{CCM}[CCM]{Chassis Control Module}
\acrodef{LCM}[LCM]{Light Control Module}
\acrodef{CANH}[CANH]{CAN high line}
\acrodef{CANL}[CANL]{CAN low line}
\acrodef{HS}[HS]{High-speed}
\acrodef{LS}[LS]{Low-Speed}
\acrodef{OBD}[OBD]{On-Board Diagnostics}
\acrodef{CMAC}[CMAC]{Cipher-based Message Authentication Code}
\acrodef{HMAC}[HMAC]{keyed-Hash Message Authentication Code}
\acrodef{MitM}[MitM]{Man in the Middle}
\acrodef{VGT}[VGT]{Variable Geometry Turbine}
\acrodef{DoS}[DoS]{Denial of Service}
\acrodef{OEM}[OEM]{Original Equipment Manufacturer}
\acrodef{OOK}[OOK]{On-Off Keying}
\acrodef{ASK}[ASK]{Amplitude-Shift Keying}
\acrodef{CRC}[CRC]{Cyclic Redundancy Check}
\acrodef{MPU}[MPU]{Micro Process Unit}
\acrodef{PWL}[PWL]{Piecewise Linear}
\acrodef{IDE}{Identifier Extension}
\acrodef{SPDT}{Single pole double throw}
\acrodef{CPSM}{Carrier Phase Shift Modulation}
\acrodef{SecOC}{Secure Onboard Communication}
\acrodef{CANsec}{CAN secure}
\acrodef{SNR}{Signal-to-Noise Ratio}
\acrodef{PSK}{Phase Shifting Keying}
\acrodef{BPSK}{Binary Phase Shifting Keying}
\acrodef{RMS}{Root Mean Square}
\acrodef{8PSK}{Eight Phase Shift Keying}
\acrodef{EMC}{Electromagnetic Compatibility}
\acrodef{NIST}{National Institute of Standards and Technology}
\acrodef{IDS}{Intrusion Detection System}
\acrodef{WCRT}{Worst-Case Response Time}
\acrodef{AUTOSAR}{Automotive Open System Architecture}
\acrodef{HSM}{Hardware Secure Module}
\acrodef{RMS}{Rate-monotonic scheduling}
\acrodef{SoC}{System-on-Chip}
\acrodef{SAE}{Society of Automotive Engineers}

\newcommand{\acintitle}[1]{\texorpdfstring{\ac{#1}}~}

\begin{document}

\title{CAN-MM: Multiplexed Message Authentication Code for Controller Area Network message authentication in road vehicles}

\author{Franco~Oberti,~\IEEEmembership{Member,~IEEE,}
        Alessandro~Savino,~\IEEEmembership{Senior~Member,~IEEE,}
        Ernesto Sanchez,~\IEEEmembership{Senior~Member,~IEEE,}
        Paolo~Casasso,
        Filippo Parisi
        and~Stefano~Di Carlo,~\IEEEmembership{Seniore~Member,~IEEE}\thanks{A. Savino, E. Sanchez and S. Di Carlo are with the Department
of Control and Computer Engineering, Politecnico di Torino, Torino, 10129, Italy. e-mail: \{alessandro.savino,ernesto.sanchez,stefano.dicarlo\}@polito.it.}\thanks{F. Oberti is with the Department
of Control and Computer Engineering, Politecnico di Torino, Torino, 10129, Italy and DUMAREY Softronix S.r.l., Torino, 10129, Italy. email: franco.oberti@dumarey.com}.
\thanks{P. Casasso and F. Parisi are with  DUMAREY Softronix S.r.l., Torino, 10129, Italy. e-mail: \{paolo.casasso,filippo.parisi\}@dumarey.com.}\thanks{This work was supported by project SERICS (PE00000014) under the MUR National Recovery and Resilience Plan funded by the European Union - NextGenerationEU}}

\maketitle

\begin{abstract}
As the automotive industry adopts more technology, the threat of cyberattacks on vehicles grows. \acp{ECU} operate in a hostile environment, raising safety concerns for drivers and passengers. Initiatives from both industry and government bodies aim to address these risks. The primary communication protocol used in the automotive industry, the standard \acp{CAN} protocol, is a target for cybercriminals due to its limitations in ensuring communication integrity. This paper proposes \ac{CAN-MM}, using frequency modulation to multiplex \ac{MAC} data with standard \ac{CAN} communication. \ac{CAN-MM} enables the transmission of \ac{MAC} payloads at reduced time cost while maintaining backward compatibility with old \ac{CAN} protocol versions. The solution is also compatible with modern evolutions of the \ac{CAN} protocol and advanced algorithms resorting to \ac{MAC} as part of the security infrastructure. 
\end{abstract}

\acresetall


\IEEEpeerreviewmaketitle

\section{Introduction}
\label{sec:CAN-MM Introduction}

Modern road vehicles, striving for improved comfort, sustainability, environmental friendliness, and safety \cite{6176573}, feature intricate onboard control systems, especially in real-time safety-critical domains \cite{6176573,8263141}. The increased interconnectivity of electronic components exacerbates this complexity. However, this sophistication also makes the automotive industry an attractive target for attackers \cite{CySTrend}, with \acp{ECU} vulnerable to cyberattacks in hostile environments \cite{8537180}.

To mitigate these risks, carmakers and governments are endorsing initiatives to bolster cybersecurity in the automotive sector (i.e., the ISO/SAE 21434:2021 standard for road vehicles cybersecurity engineering \cite{iso21434}, and the ISO/PAS 5112:2022 guidelines for auditing cybersecurity engineering \cite{iso5112}). Additionally, the \ac{UNECE} has introduced new regulations for vehicle cybersecurity and software updates, delivered through the WP.29 package \cite{unece-155-2021,unece-156-2021}. The automotive industry is working harder to make their products more secure and to research ways to address serious security threats that take advantage of communication between modules \cite{myart,9525579, LINMM}.

The \ac{CAN} protocol is central to automotive communication. Therefore, ensuring robust security measures within \ac{CAN} communication is crucial to uphold the integrity and safety of modern vehicles \cite{6542519}. Detailed insights into potential \ac{CAN} threats and related countermeasures are provided in \cite{ROMP1,9439954,8697772}. Country-specific regulations mandate specific \ac{CAN} messages accessible through an \ac{OBD} port in every vehicle \cite{EPA,UNECE}. Ensuring the integrity (i.e., immunity to tampering) and authenticity (i.e., originating from an authorized source) of \ac{CAN} messages is, therefore, critical to prevent unauthorized access and ensure the safety and operational efficiency of essential functionalities of the vehicle \cite{articlemac1,articlemac2,articlemac3}. To achieve this, the \ac{SecOC} and Crypto Stack defined in \ac{AUTOSAR} require the incorporation of a \ac{MAC} digest within the payload of each data frame \cite{autos}. However, integrating a \ac{MAC} digest in a \ac{CAN} frame presents compatibility issues, feasible only for specific \ac{CAN} protocol versions and resulting in back-compatibility challenges~\cite{10075498}.

This paper proposes a technique named \ac{CAN-MM}, offering a novel approach to \ac{MAC} transmission. This technique enables the multiplexing of the \ac{MAC} alongside data transmission without altering the original frame format, ensuring full compatibility with all versions of the standard \ac{CAN} protocol. The main objective of \ac{CAN-MM} technology is to integrate a \ac{SoC} compatible \ac{MAC} in the \ac{CAN} version 2.0 to enable achieving a security level that matches the most recent advancements, such as \ac{SecOC} utilizing \ac{MAC} with \ac{CAN-FD}. Moreover, this approach addresses the authentication timing challenges identified by Ikumapayi et al.\cite{10075498}. Eventually, by freeing data bytes from the \ac{CAN} frame, it offers a novel approach to incorporate the \ac{MAC} in signature schemas, authentication protocols, or key exchange mechanisms, such as \cite{Groza:2017aa}.

The article is organized as follows: \autoref{sec:CAN-overview} gives some background on the \ac{CAN} network, including vulnerabilities and common attacks, while \autoref{sec:related_works} reports the state-of-the-art literature on \ac{CAN} security. Section \autoref{sec:CANMM_TA} describes the \ac{CAN-MM} architecture. Section \ref{sec:results} provides experimental results, and  \autoref{sec:conclusions} summarizes the main contributions and concludes the paper.

\section{Background}
\label{sec:CAN-overview}

On-board \acp{ECU} play a crucial role in automotive applications by managing subsystems and facilitating real-time communication with sensors and actuators \cite{albert2004comparison}. The \ac{CAN} bus, a primary vehicle network, adheres to safety guidelines, ensuring reliable communication in noisy environments. The \ac{CAN} electrical signal, transmitted differentially through \ac{CANH} and \ac{CANL}, minimizes noise impact from motors, ignition systems, and switching contacts. \ac{HS} (ISO 11898-2 \cite{iso118982}) and \ac{LS} (ISO 11898-3 \cite{iso118983}) interfaces provide varying throughput capabilities based on different voltage levels. In \ac{HS} \ac{CAN}, dominant bit transmission (logic 0) raises \ac{CANH} to 3.5V and lowers \ac{CANL} to 1.5V, creating a 2V voltage difference. Recessive bit transmission (logic 1) maintains both \ac{CANH} and \ac{CANL} at 2.5V with minimal voltage difference. A differential voltage above 0.9V indicates a dominant level (logic 0), while below 0.5V denotes a recessive level (logic 1), ensuring reliable communication in noisy environments. Twisted-pair conductors are commonly used for physical transmission lines to mitigate magnetic interference. 

Multiple \ac{CAN} protocol variants exist, each supporting different transmission speeds and frame payload sizes. \ac{CAN-FD} and \ac{CAN} 2.0 protocols differ in maximum transmission speed and payload size, with \ac{CAN} 2.0 limited to 8 bytes and \ac{CAN-FD} extending to 64 bytes. Despite \ac{CAN-FD} supporting larger payloads, many applications still use 8-byte payloads to ensure compatibility with existing vehicle \ac{CAN} database~\cite{Kaiser:2019aa,Bi:2023aa,Gazdag:2023aa}. \ac{CAN-XL}, a newer version meeting ISO/TC 22/SC 31 Data communication standards \cite{iso22sc31}, offers features like extended data payload capacity (up to 2,048 bytes) and higher communication speeds ranging from 500 kbit/s to 5 Mbit/s, with potential speeds reaching 12 Mbit/s in the \ac{CAN} SIC XL FAST configuration. The \ac{CAN} SIC XL FAST baud rate is comparable to the 10BASE-T1S technology, also known as Vehicle Ethernet, providing 10 Mbit/s bandwidth over a single-pair physical layer.
The original \ac{CAN} protocol includes no built-in security features. Additionally, country-based regulations require the provision of an \ac{OBD} port \cite{EPA,UNECE}, commonly located within vehicles, enabling access to legislative diagnostic messages. These messages, transmitted in plaintext to comply with legislative mandates, introduce considerable security vulnerabilities. 

In an endeavor to mitigate these risks, the \ac{SecOC} framework, explicitly designed for \ac{CAN-FD}, along with \ac{CANsec} for \ac{CAN-XL}, has been promulgated. These methodologies expressly elevate the principles of data integrity and authenticity over confidentiality \cite{CIAtriad}. The critical role played by the \ac{CAN} bus in the domain of automotive communications mandates a comprehensive investigation into its security weaknesses, potential avenues for attack, and the methods by which such attacks may be carried out \cite{Bozdal:2018aa, ROMP3, Song:2016aa}.

The attack surface of a \ac{CAN} presents numerous potential vulnerabilities attackers could exploit. This encompasses strategies for unauthorized access, undermining data integrity, data breaches, executing hijacking maneuvers, or hindering the system. Despite the variety of attack vectors against \ac{CAN} networks, two main types of attacks have been reported in the literature: (i) \ac{MitM}~\cite{Gazdag:2021aa} and (ii) Replay Attacks~\cite{Noureldeen:2017aa}.

\begin{figure}[htb]
    \centering
    \includegraphics[width=0.98\columnwidth]{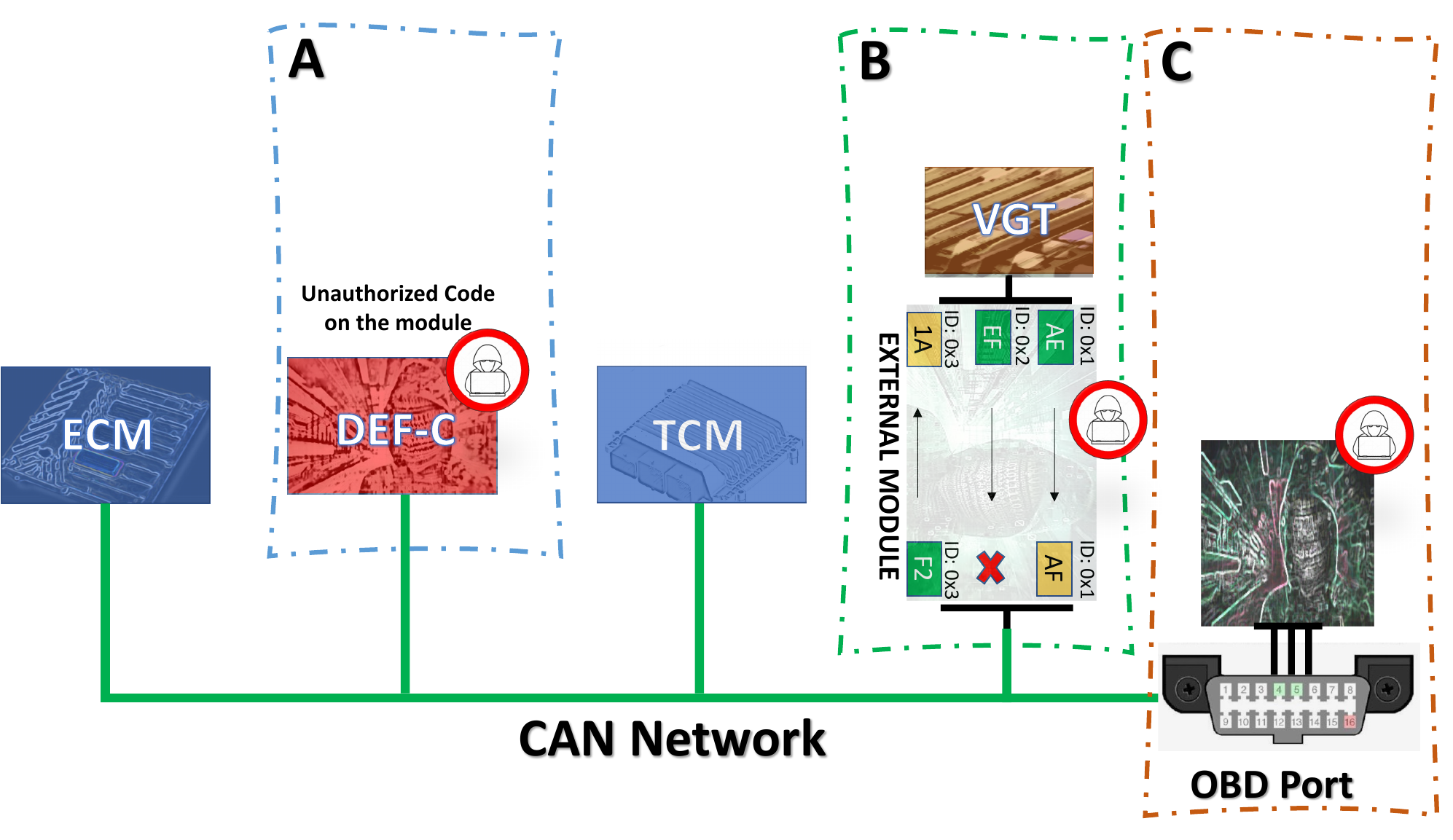}
    \caption{Vehicle \ac{CAN} network surface attack scheme. A small \ac{CAN} vehicle network scheme composed of 4 modules: \ac{ECM}, \ac{TCM}, \ac{DEFC}, and \ac{VGT}. These \acp{ECU} communicate with sensors and actuators in real-time, making integration essential for their operation. (A) Corrupted vehicle \ac{CAN} node runs unauthorized code. (B) Attack vector through external \ac{CAN} module plugged upstream to \ac{CAN} victim node. (C) The external \ac{CAN} module directly accesses the \ac{OBD} port inside the vehicle cabin. }
    \label{fig:surfacescheme}
\end{figure}

Figure \ref{fig:surfacescheme} illustrates three prevalent automotive attack settings that target the \ac{CAN} protocol. Each setting is effectively utilized in \ac{MitM} and Replay Attacks. Figure \ref{fig:surfacescheme}-A demonstrates an attack through a compromised \ac{CAN} node, where unauthorized software takes control. This can occur via the corruption of the \ac{CAN} controller's firmware or by exploiting software module vulnerabilities, such as a buffer overflow. In \autoref{fig:surfacescheme}-B, an attack is facilitated by a hardware module that isolates the victim node from the rest of the vehicle network, enabling the interception and manipulation of \ac{CAN} traffic. The final scheme, depicted in Figure \ref{fig:surfacescheme}-C, involves connecting an external module to the vehicle's \ac{OBD} port, granting direct access to the \ac{CAN} bus. Various commercially available, low-cost \ac{CAN} modules that feature Bluetooth connectivity support this approach, allowing for programmability via mobile applications. These settings are crucial in laying the groundwork for advanced \ac{CAN} attacks, exemplified by the Janus Attack \cite{Tindell:2024aa} and the Cloak Attack \cite{Yue:2021aa}.

The Janus Attack, a new and sophisticated threat in \ac{CAN} protocol~\cite{Tindell:2024aa}, leverages the CAN protocol synchronization rules and targets devices with different sample points. It involves transmitting a single \ac{CAN} frame with dual payloads, causing targeted devices to interpret divergent data compared to others in the network. This undermines the atomic multicast principle of \ac{CAN}, critical for system integrity. It operates by coercing all \ac{CAN} controllers to synchronize simultaneously, then manipulating the \ac{CAN} bus level after the first one has sampled the bus but before another does, resulting in valid frames with differing payloads as it exploits the characteristics of the two different payloads to have the same size. 

A cloak attack in cybersecurity involves manipulating bit signals to deceive networked \acp{ECU}~\cite{Yue:2021aa}. The main idea is that the attacker leverages the different sampling times of two receivers to craft two different frames (FrameA and FrameB). The difference is represented by a selection of bits the attacker alters after the first receiver samples the frame (FrameA). Appropriately crafted, the bit-changes in the second frame (FrameB) can avoid triggering re-synchronization mechanisms, aiming for an optimized bit-string with minimal detection and errors in the \ac{CRC} field (as the \ac{CRC} code will be based on the original content of FrameA). If the attacker achieves such duplication, it can generate out-of-sync data in \acp{ECU}.

The Replay Attack shares similarities with \ac{MitM} attack. To execute this attack, the attacker must perform a learning phase by monitoring the network and collecting a certain amount of \ac{CAN} frames. Later, the attacker replays these previously collected frames on the network to achieve a target behavior. Unfortunately, this attack does not require the attacker to possess specific skills, expertise, or advanced knowledge about vehicle \ac{CAN} networks.

These clusters of attacks can be successfully mitigated by linking a \ac{CAN} Frame payload to a unique \ac{MAC} that is directly derived from the frame data. Yet, the \ac{MAC} alone is insufficient for replay attacks due to the \ac{CAN} payload with identical data producing the same digest. Hence, adopting a rolling counter tied to the data is advised to achieve different digests while maintaining data parity.

The \ac{MAC} effectively mitigates threats but may also introduce weaknesses in the framework system. This is especially significant in safety-critical, hard real-time systems like \ac{ECM}, \ac{TCM}, etc. Ikumapayiet al. formalizes the impact that authentication schemes have on the real-time performance of messages over \ac{CAN}, \ac{CAN-FD}, and \ac{CAN-XL} based on response time analysis. A \ac{CAN} frame is schedulable if its \ac{WCRT} is less than or equal to its deadline. Message deadlines may be implicit, i.e., equal to their period, or explicit (constrained). In particular, Ikumapayi et al. \cite{10075498} demonstrated that adding a \ac{MAC} to the payload of \ac{CAN}, \ac{CAN-FD}, and \ac{CAN-XL} messages might impact the schedulability and the meeting of deadlines based on the percentage of utilization. In particular, on classical \ac{CAN}, after 70\% of utilization, almost all messages fail to meet the deadlines. On the other hand, \ac{CAN-FD} and \ac{CAN-XL} exhibit higher schedulable resilience (it drops when the percentage of bus utilization rises to 80-90\%) thanks to the faster bit rate. Nevertheless, pushing such high bus utilization can be malevolent.

When the \ac{CAN} frames include the \ac{MAC} in their payload before utilizing the data, the \ac{MAC} shall be verified as a success. Modern \ac{ECU}s are generally equipped with a \ac{HSM}, a dedicated \ac{SoC} module that manages all cryptographic and security functions, including verifying MACs. The host system is momentarily suspended during the verification process by the \ac{HSM}. In the context of real-time systems, an attacker might take advantage of this by injecting or flooding the \ac{CAN} vehicle network with secure \ac{CAN} frames that possess a legitimate ID but include counterfeit data and \ac{MAC}. This situation leads to the \ac{HSM} being overwhelmed with \ac{MAC} verification requests that fail, while the host system is forced into repeated waiting periods, causing abnormal delays \cite{9106836}. These delays can significantly disrupt the system’s capacity to adhere to its real-time deadlines, necessitating the initiation of safety system recoveries to address the failure to meet these critical timing constraints.
\section{Related Works}
\label{sec:related_works}

As the original version of \ac{CAN} protocol did not include any security support, researchers have come a long way to support it on top of the existing infrastructure or by proposing enhanced versions.

First attempts to improve the security of the \ac{CAN} protocol and improve resistance to attacks involved including a \ac{MAC} digest for integrity and authenticity assurance \cite{4453831}, often employing \ac{CMAC} or \ac{HMAC} signatures, depending on hardware support. The CAN+ protocol, introduced by Ziermann et al. in \cite{CAN+}, aimed to enhance \ac{CAN} data rates by relaxing constraints during specific transmission time slots. While the \ac{CAN} application can benefit from the increased speed, its assessment lacked consideration for \ac{EMC} and disturbance handling, which is crucial in the automotive domain. Furthermore, CAN+ relies on media access characteristics not present in the latest \ac{CAN-FD} and \ac{CAN-XL} protocols, which offer higher payload sizes and data rates. Despite advancements, minimizing latency in \ac{MAC} signature reception and checking remains essential in \ac{CAN-FD} and \ac{CAN-XL}, which offer increased payload size and data rates.

Significant advancements have been made to enhance broadcast authentication mechanisms, capitalizing on the increased data rate of CAN+. Van Herreveg et al. introduced CanAuth \cite{van2011canauth}, a backward-compatible broadcast message authentication protocol for the \ac{CAN} bus. This protocol meticulously follows \ac{CAN} specifications, prioritizing ID-oriented authentication while addressing authentication delays and time synchronization concerns. However, Groza et al. \cite{Groza:2017aa} point out that CanAuth's drawback lies in managing many keys associated with message IDs, raising security concerns. In response, they propose the LIBrA-CAN protocol as an alternative. Both LiBrA-CAN and CanAuth share the goal of enhancing \ac{CAN} communication security but adopt distinct approaches and mechanisms. LiBrA-CAN emphasizes decentralized broadcast-based arbitration and lightweight implementation, ensuring resilience against replay attacks and flexibility in configuration. On the other hand, CanAuth focuses on message authentication and verification, providing robust protection against unauthorized access and tampering. To preserve the integrity of the physical layer, Hazem et al. \cite{hazem2012lcap} put forth LCAP, a Lightweight CAN Authentication Protocol for Securing In-Vehicle Networks.

All previous works point out that the \ac{MAC} size can significantly impact the resistance to attacks, i.e., the \ac{MAC} size and the time required to elaborate it. To tackle the time constraints, authors in \cite{Schmandt:2017aa} proposed a truncated \ac{MAC}, justified by the average data size of 15,768 messages from a 2010 Toyota Prius during a 12.27-minute use case. They noted that only a part of the 8 bytes available in the \ac{CAN} frames were used, making room for a short \ac{MAC}. Following a similar direction, to further reduce the schema complexity and support all possible \ac{CAN} protocols, very recently, Luo et al. \cite{Luo:2021aa} proposed a lightweight schema based on the introduction of the \ac{MAC} in place of the \ac{CRC} field in the 2.0 version of the protocol. While the authors demonstrated the capability of their approach, the back compatibility with standard hardware is not guaranteed, as they will check a \ac{CRC} value that is not correct.

In general, both approaches go against \ac{NIST} guidelines, stating that a truncated \ac{MAC} digest below 4 bytes compromises cyber resilience \cite{rfc4493}. Ikumapayi et al. \cite{10075498} have explored the impact of adding authentication codes as separate messages, noting potential strain on timely delivery, especially given size constraints. As the authors noted, the effect of reserving more than four bytes in \ac{CAN} 2.0 (i.e., 24Bit-CMAC-8Bit-FV) limits data interchangeability as it requires adding an extra frame to contain the remaining bytes that do not fit into the original frame. However, secure \ac{CAN-FD} and \ac{CAN-XL} protocols support \ac{MAC} digest sizes from 4 to 16 bytes, accommodating complex protocols like authentications as demonstrated by \cite{Groza:2017aa}. 
Yet, upgrading an entire vehicle network to these protocols involves benefits and extra costs~\cite{Esparza:2015aa}, which are left to the manufacturer to evaluate.

Eventually, it is worth mentioning that some recent works support authentication and confidentiality without resorting to \ac{MAC}~\cite{Pese:2021aa}. They include only cryptography techniques in the handshake phase, leading to a tiny increase in the latency, limited to hundreds of \textmu s, paying with reduced security if compared with schemas resorting to \ac{MAC}~\cite{Radu:2016aa, Sugashima:2016aa, Groza:2017aa}.

Modulation techniques are not new in the security of the \ac{CAN} protocol; recent efforts by Michaels et al. \cite{ROMP2} introduced modulation techniques to enhance the security of the \ac{CAN} protocol. Their proposal incorporates a rolling secret (watermark) aligned with primary bus messages through multiplexing based on \ac{BPSK} modulation. While this multiplexed watermark significantly improves security by ensuring transmitted message authentication, it solely addresses this aspect, leaving incomplete coverage to attacks such as \ac{MitM}, as the watermark can be forged. 

\section{\acintitle{CAN-MM}~~Technology}
\label{sec:CANMM_TA}

\ac{CAN-MM} technology offers a non-intrusive solution for implementing \ac{MAC}-based message authentication and integrity checks without compromising payload capacity or backward compatibility with all \ac{CAN} standard versions. This approach is especially relevant for \ac{CAN} 2.0 applications, enabling the development of a secure \ac{CAN} network with an large \ac{MAC} digest size. Additionally, \ac{CAN-MM} enhances response time and performance of \ac{MAC} digest computation across all \ac{CAN} versions.

Essentially, the underlying concept of \ac{CAN-MM} involves utilizing digital modulation techniques (i.e., \ac{OOK}) to multiplex the transmission of the \ac{MAC} digest with the original \ac{CAN} frame payload. The \ac{OOK} is a simple digital modulation scheme based on \ac{ASK} commonly used in telecommunication \cite{onoffkeing,ASK}. \ac{OOK} transmits a logical one by sending a carrier wave signal, while the absence of the carrier wave represents a logical zero.

The \ac{MAC} information is encoded by switching the carrier wave on and off. A logical zero is transmitted on the bus by generating the original CAN signals, while in the case of a logical one, a wave is added to the standard CAN electric signals (in both \ac{CANH} and \ac{CANL}). This wave acts as a carrier. Its amplitude is a configured parameter, with a value of $Vpp=300mV$ in this study, to ensure sufficient margins when reconstructing the original signal at the receiver's side. 

\begin{figure*}[hbt]
    \centering
    \includegraphics[width=0.94\textwidth]{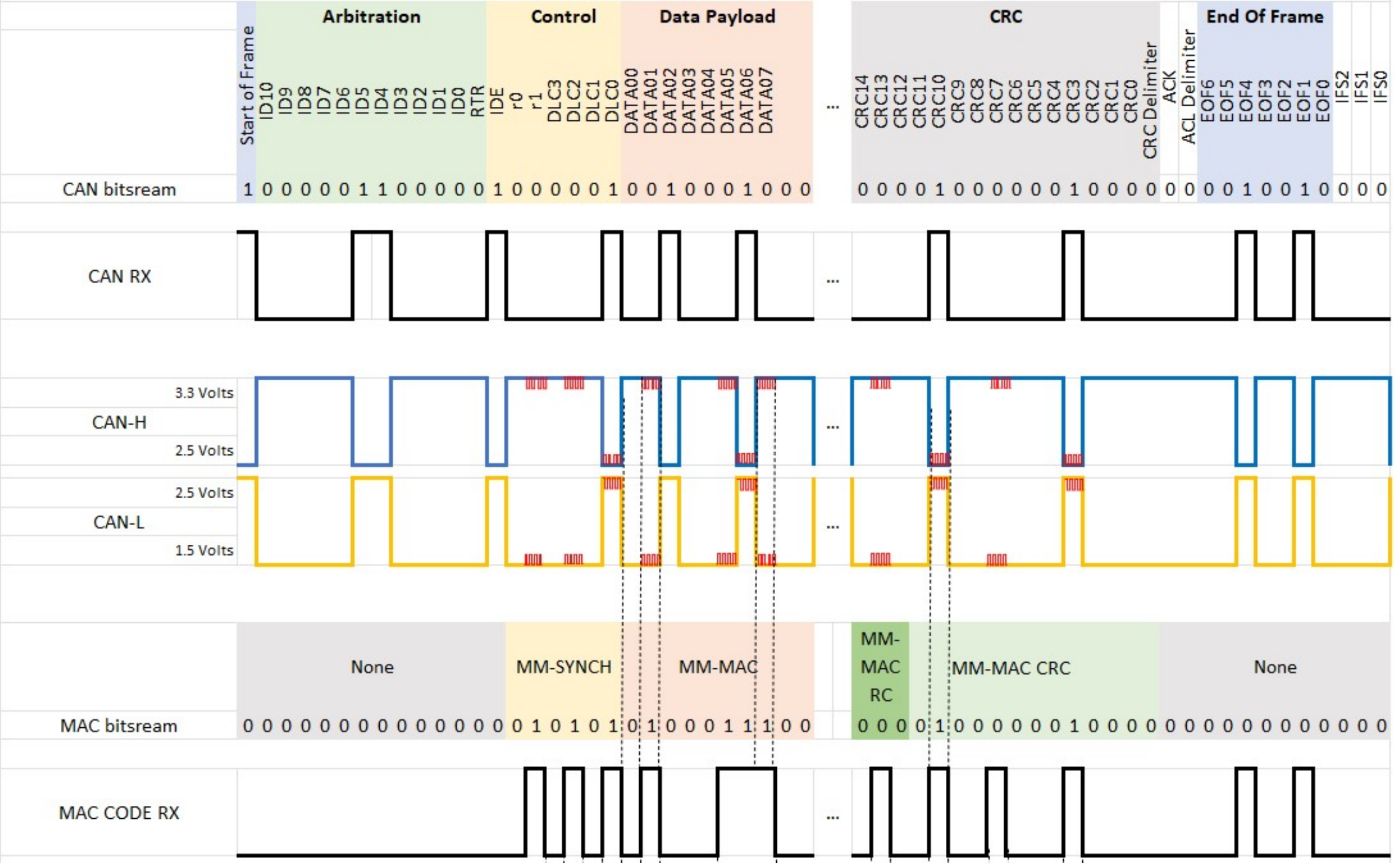}
    \caption{Physical Electrical CAN-MM Signal Scheme}
    \label{fig:CNMMSCH}
 \end{figure*}

To combine the signals from the \ac{CAN} frame and \ac{MAC} digest, the \ac{CAN-MM} system necessitates appropriate synchronization, as depicted in \autoref{fig:CNMMSCH}. The \ac{IDE} bit of the \ac{CAN} Control field initiates the synchronization procedure. During this procedure, a synchronization sequence of logic "1" and "0" is introduced on the \ac{MAC} CODE RX line for the entire duration of the Control field. These values are modulated with the content of the Control field. Subsequently, the \ac{MAC} digest is modulated onto the data payload. Finally, to enhance the reliability of the system, the \ac{CRC} of the \ac{MAC} digest is modulated onto the \ac{CRC} slot of the payload. The \ac{CRC} is a specialized checker to detect transmission errors. Multiplexing the \ac{MAC} digest directly with the message ensures a strong link between the \ac{MAC} code and the corresponding message, bolstering security by minimizing vulnerabilities such as message and code separation.

\begin{figure}[htb]
    \centering
    \includegraphics[width=0.98\columnwidth]{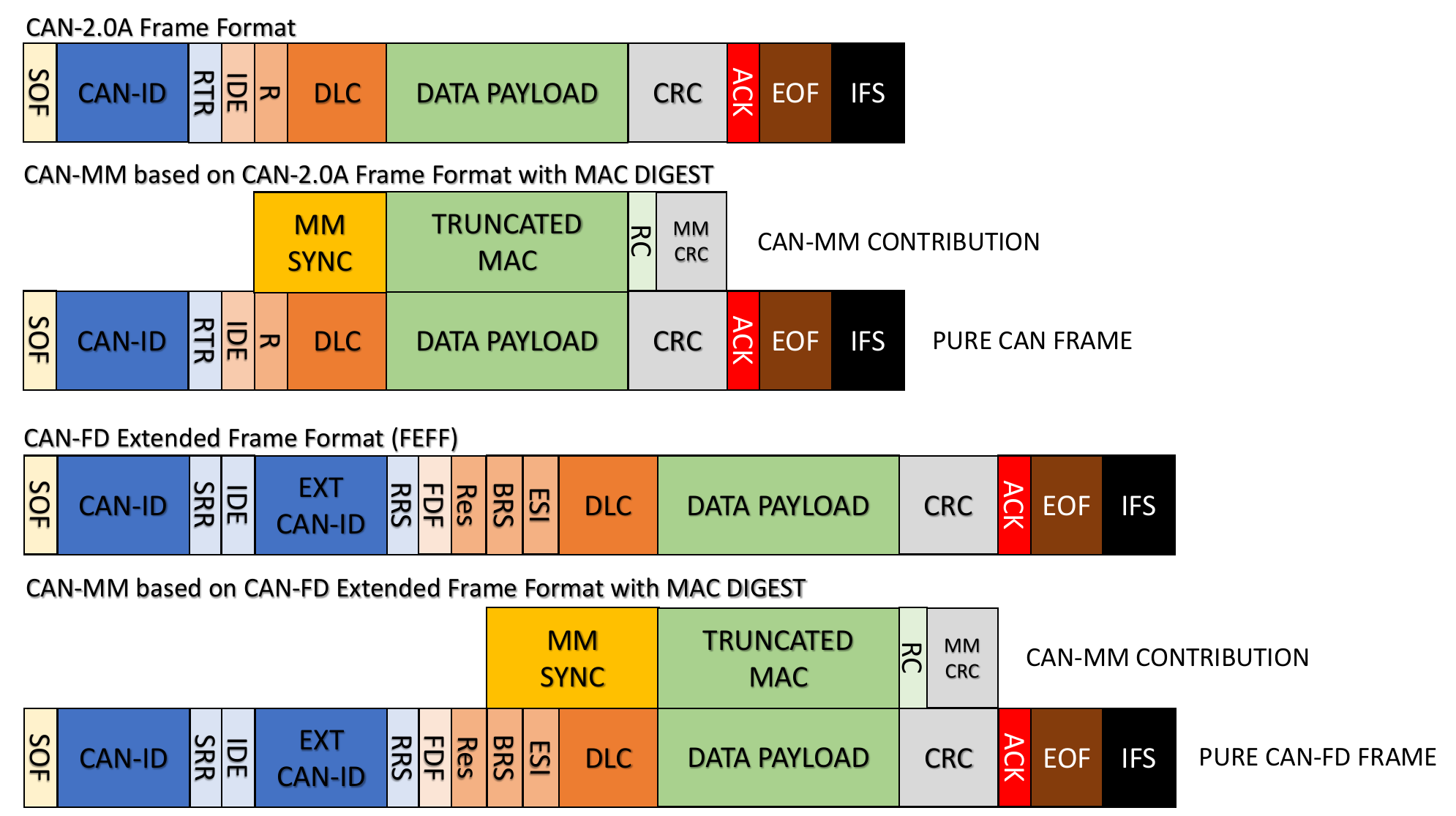}
    \caption{Application of the CAN-MM technology to both \ac{CAN} 2.0 and \ac{CAN-FD} frames}
    \label{fig:CANMMVSCANFD}
  \end{figure}

\autoref{fig:CANMMVSCANFD} depicts the effect of using the \ac{CAN-MM} approach on the \ac{CAN} 2.0 and \ac{CAN-FD} frames. In both cases, modulating the \ac{MAC} helps maintain the full payload capacity of the frame; when the frame is long enough, e.g., the \ac{CAN-FD}, it reduces the necessary size of the frame while retaining the same amount of information. This reduction limits the need for the extra transmission time caused by appending the \ac{MAC} to the data payload or as extra frames~\cite{10075498} when the selected \ac{MAC} length is above 64 bits. It also might help optimize the system's real-time performance and the \ac{CAN} bus load of the entire vehicle network.

The \ac{CAN-MM} architecture consists of two main blocks: a transmitter and a receiver module.

\begin{figure*}[htb]
    \centering
    \includegraphics[width=0.8\textwidth]{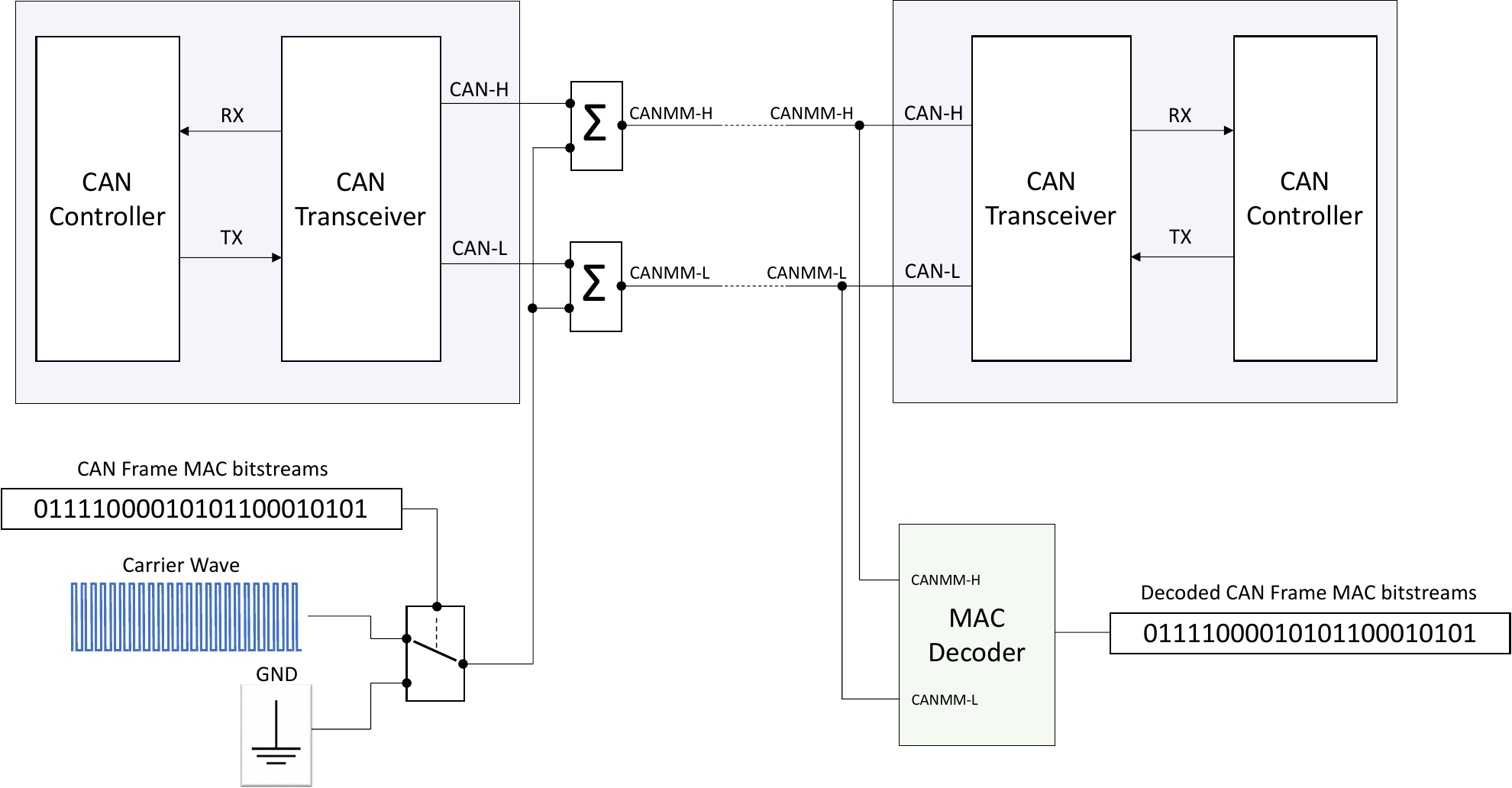}
    \caption{CAN-MM Transmitter \& Receiver block scheme}
    \label{fig:7}
 \end{figure*}

In the left part of \autoref{fig:7}, the original transmitter (\ac{CAN} controller and \ac{CAN} transceiver blocks) is coupled with the additional functional components required to implement the \ac{CAN-MM} schema in the bottom left. A multiplexer block is employed to multiplex the \ac{MAC}-related information. This block includes a diverter switch \cite{6573413} with two inputs, namely a carrier supplied by an internal generator and ground. The modulated \ac{CAN} signal is applied to both \ac{CANH} and \ac{CANL}. The multiplexer is controlled by the \ac{MAC} bitstream to provide a carrier as output when the corresponding \ac{MAC} bit is one and no contribution when the corresponding bit of the \ac{MAC} is zero. The multiplexer control line is synchronized with the \ac{CAN} controller to multiplex the \ac{MAC} information with the \ac{CAN} payload.
 
In the right part of \autoref{fig:7}, the receiver includes a decoder block that is responsible for extracting the multiplexed \ac{MAC} bitstream sequence from the payload of the \ac{CAN} frame. The \ac{MAC} decoder block utilizes a hybrid analog-digital electronic network to extract the correct \ac{CAN-MM} contributions encapsulated in the \ac{CAN} physical signal. The standard \ac{CAN} receiver and the \ac{MAC} decoder operate in parallel, eliminating the necessity of extra computation time for MAC extraction. While the transceiver processes the \ac{CAN} frame, the decoder reconstructs the \ac{MAC} bitstream. This allows the \ac{ECU} to receive the \ac{CAN} data payload and its \ac{MAC} code in a shorter time window compared to existing solutions \cite{6542519}.

The custom \ac{CAN-MM} components are situated downstream of the standard \ac{CAN} interface to ensure full electrical compatibility with existing \ac{CAN} interfaces.

\begin{figure}[hbt]
    \centering
    \includegraphics[width=0.98\columnwidth]{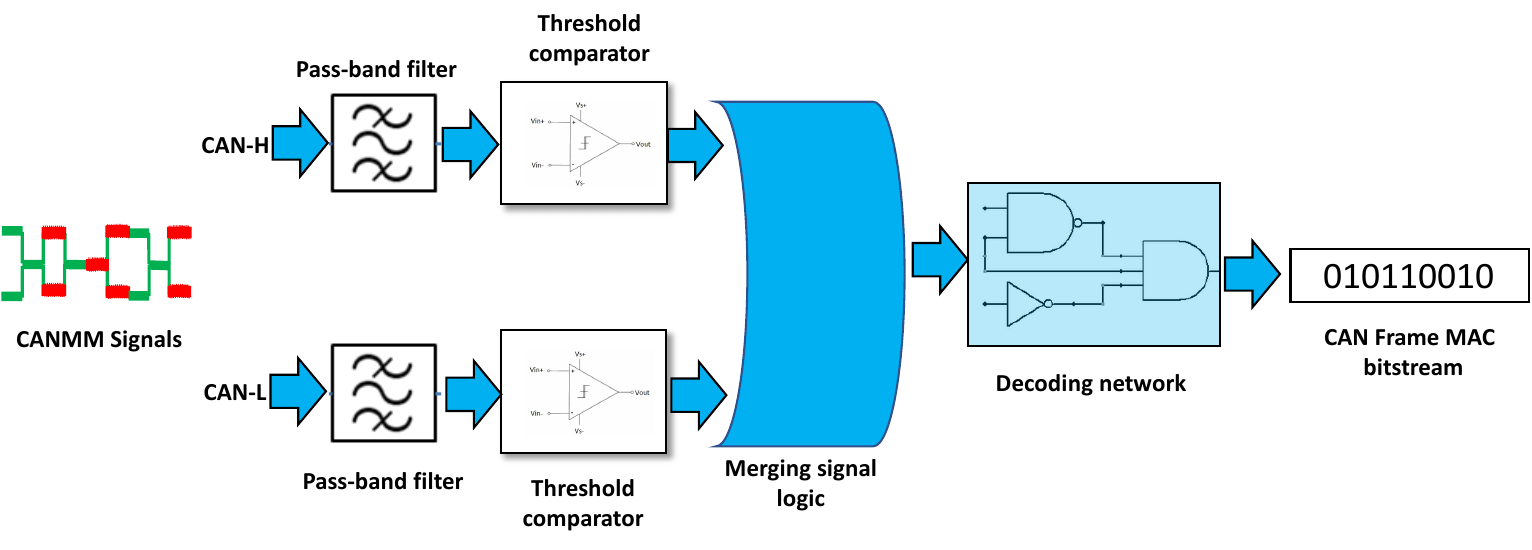}
    \caption{CAN-MM MAC decoder Type-A block scheme}
    \label{fig:9}
 \end{figure}

To further explain the \ac{CAN-MM} decoder, \autoref{fig:9} shows the complexity of the analog-digital electronic required. The decoder is composed of four stages, which are as follows:

\begin{itemize}
\item \emph{Filtering}: This stage is replicated for both \ac{CANH} and \ac{CANL}. It includes a band-pass filter with a center frequency \emph{fc} at the frequency of the carrier signal.
\item \emph{Comparing}: This stage is a threshold comparator that operates on both \ac{CANH} and \ac{CANL} to identify the specific area where the carrier signal is present.
\item \emph{Conjunction}: This stage combines the analog data from both \ac{CAN} lines into a single digital signal stream.
\item \emph{Counter}: The final stage is a logical network that identifies the area of the carrier signal in the digital domain.
\end{itemize}

These stages work together in a highly coordinated fashion to accurately extract the modulated information from the \ac{CAN} channel.

\section{Validation Model}
\label{sec:MdlVld}

\subsection{Experimental setup}

\begin{figure}[htb]
\centering
\includegraphics[width=0.98\columnwidth]{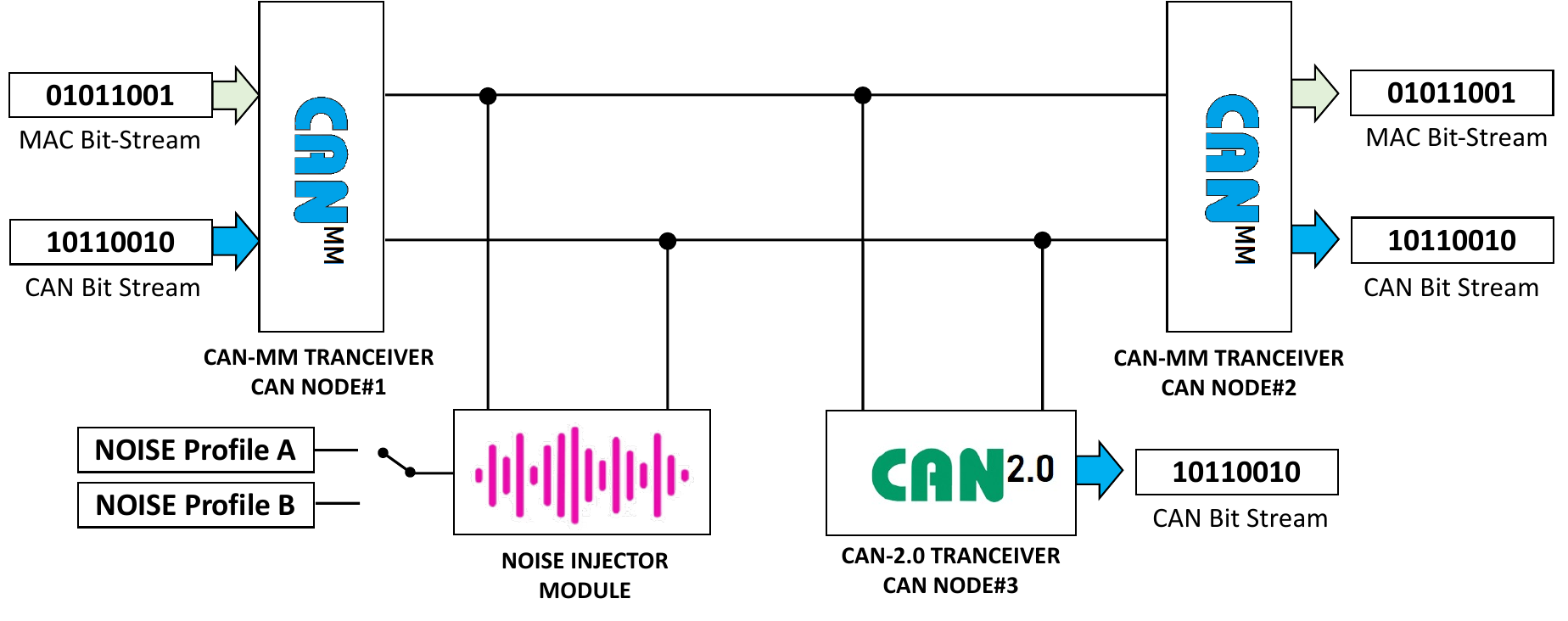}
\caption{Block scheme of the CAN-MM validation setup}
\label{fig:14}
\end{figure}

The validation of the \ac{CAN-MM} architecture considers a typical application scenario, specifically, a standard automotive \ac{CAN} 2.0 network operating at a speed of 500kbps. A hybrid automotive \ac{CAN} network comprising three \ac{CAN} nodes was designed and simulated using the LTSpice~\cite{ltspice} simulation environment to validate the architecture. Two nodes were \ac{CAN-MM} transceivers, one serving as a transmitter and the other as a receiver. The third node was a standard \ac{CAN} version 2.0 receiver. This setup enabled the validation and verification of the \ac{CAN-MM} functionality and its backward compatibility with standard \ac{CAN} transceivers. The complete block diagram for this configuration is presented in \autoref{fig:14}.

\subsection{Noise and interference analysis \label{sec:noismodel}}

\ac{CAN} systems boast a robust immunity to ground noise and electromagnetic interference, thanks to differentially transmitted information, independent ground reference, usage of twisted-pair cabling, and balanced differential transceivers.  

\begin{figure}[htb]
\centering
\includegraphics[width=0.98\columnwidth]{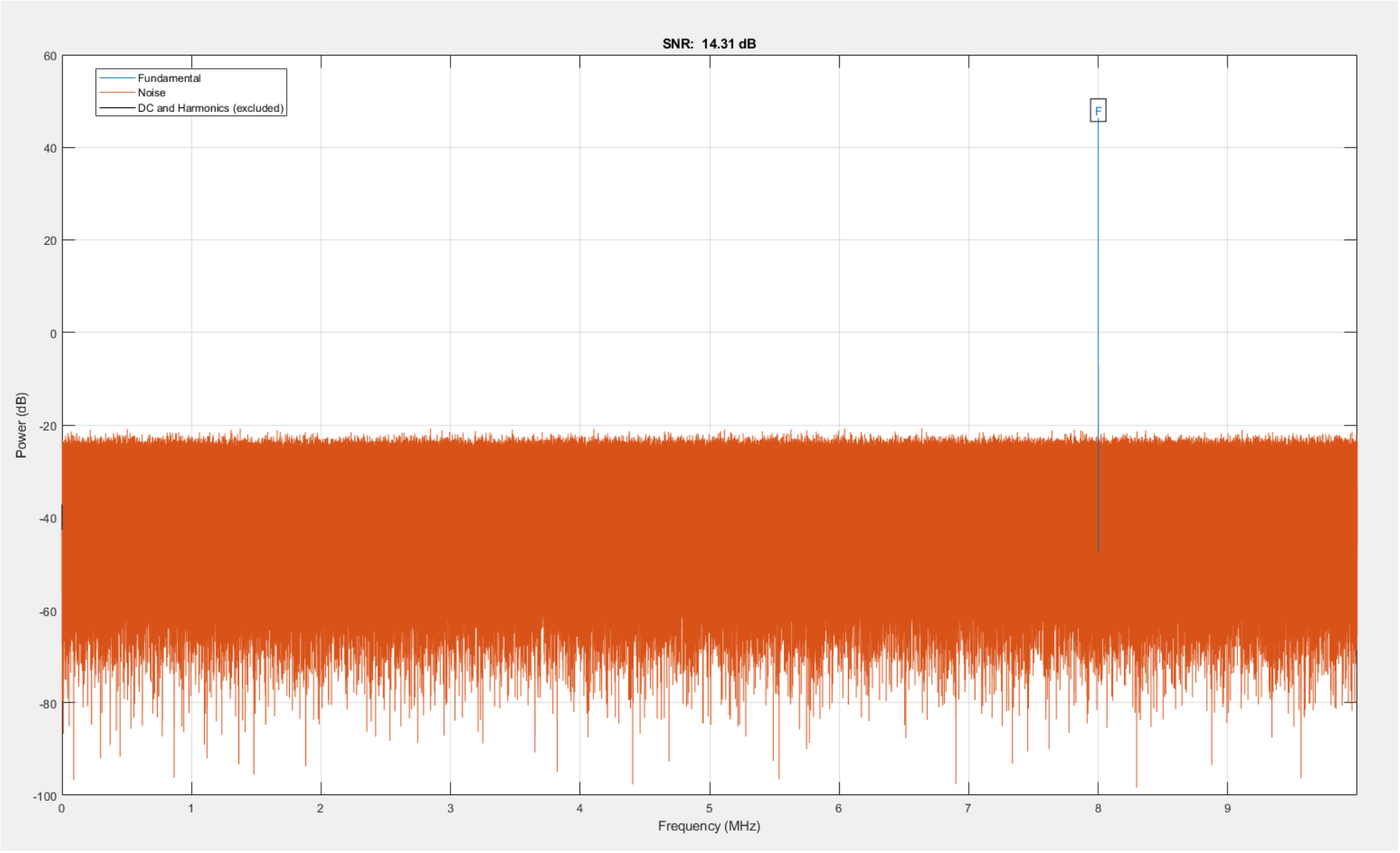}
\caption{SNR graph for real CAN recorded signals}
\label{fig:SNR}
\end{figure}

Since the \ac{CAN-MM} technology is modifying the original profile of the \ac{CAN} signals, evaluating it under realistic noisy environments is crucial. A validation environment simulated standard vehicle noise to assess noise and interference effects on \ac{CAN-MM} technology. The noise profile is acquired using a multi-protocol vehicle interface device connected to an actual vehicle's \ac{OBD} port. The device, programmed to transmit a specific \ac{CAN} frame to the \ac{ECM}, captures the physical \ac{CAN} signal via an oscilloscope. Direct access to the \ac{CAN} bus input of the \ac{ECU} is facilitated through a break-out box. The noise profile is obtained during engine idle, aligned with specifications from various research papers \cite{7522841,8549084,5982893}. Noise signals, recorded from both \ac{CAN} lines with the same phase, cover frequencies from 10kHz to 10MHz, with amplitudes between -100mV and 100mV. \ac{SNR} calculations involve computations on two identical carrier signals with a peak-to-peak amplitude of 300mV. The \ac{SNR} for this scenario was calculated to be approximately 14.31 dB (\autoref{fig:SNR}). This value provides insight into the signal's quality relative to the background noise with the current parameters.

\subsection{SPICE Model}

The SPICE simulation incorporates input signals, such as the \ac{CAN} bitstream and its associated \ac{MAC}, generated from \ac{PWL} files. Supplementary signals, including noise profiles, follow the same method with their respective \ac{PWL} files. Standard library parts provided by the tool are utilized for the remaining design components. 

\begin{figure}[htb]
    \centering
    \includegraphics[width=0.98\columnwidth]{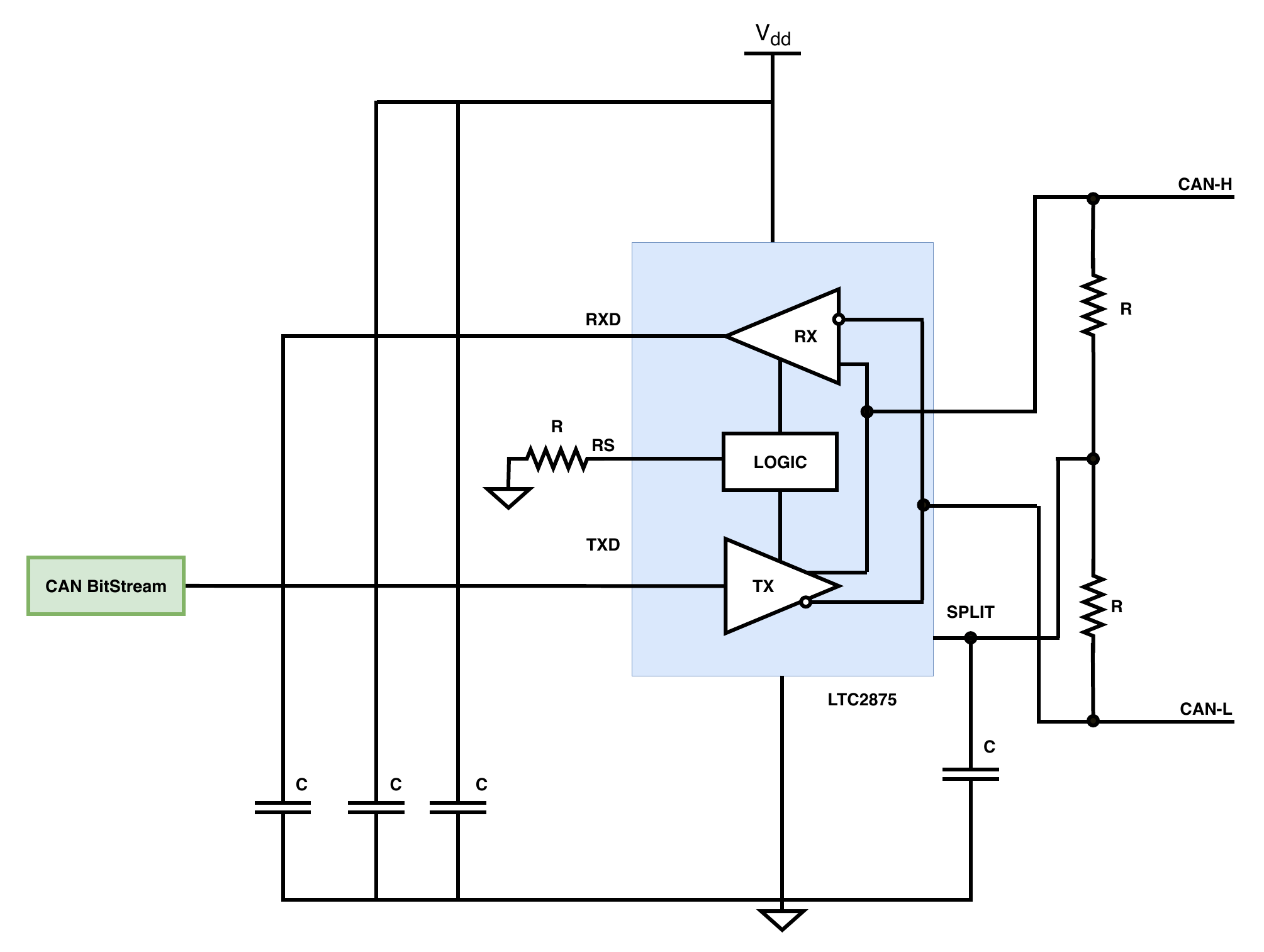}
\caption{CAN-MM Transceiver - Stage 1 - SPICE Block }
\label{CAN_module}
\end{figure}

\begin{figure}[htb]
\centering
\includegraphics[width=0.98\columnwidth]{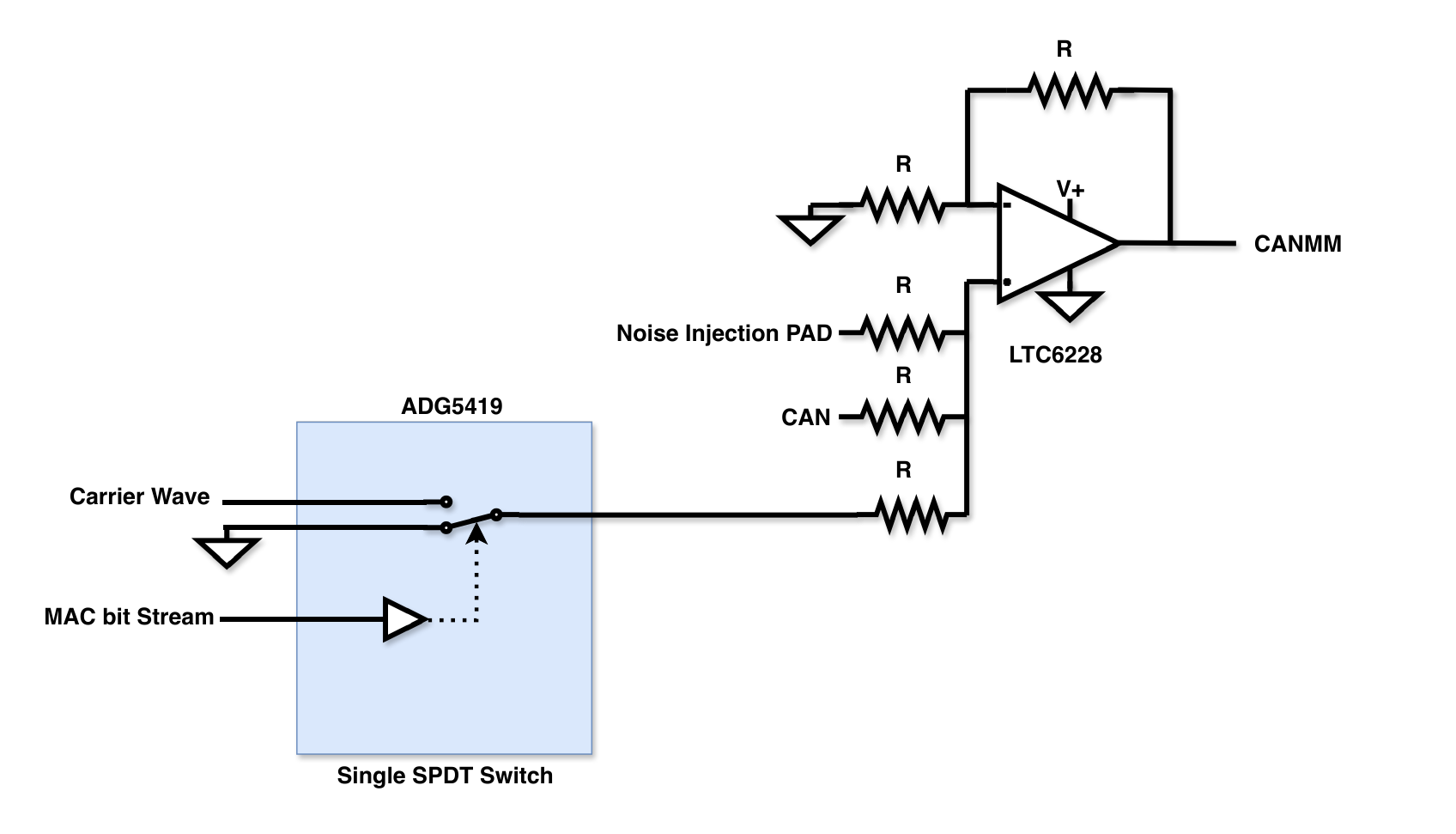}
\caption{CAN-MM Transceiver - Stage 2 - SPICE Block}
\label{CANMMT}
\end{figure}

In this setup, the LTC2875 standard \ac{CAN} transceiver (refer to \autoref{CAN_module}) is employed, as depicted in \autoref{fig:14}. The \ac{CAN-MM} added part features a High Voltage Latch-Up Proof and a \ac{SPDT} Switch. Depending on the control value, this block outputs either the carrier wave or zero, subsequently added to the \ac{CANH} and \ac{CANL} signals provided by LTC2875, along with the noise contributions (see \autoref{CANMMT}).

\begin{figure}[htb]
\centering
\includegraphics[width=0.98\columnwidth]{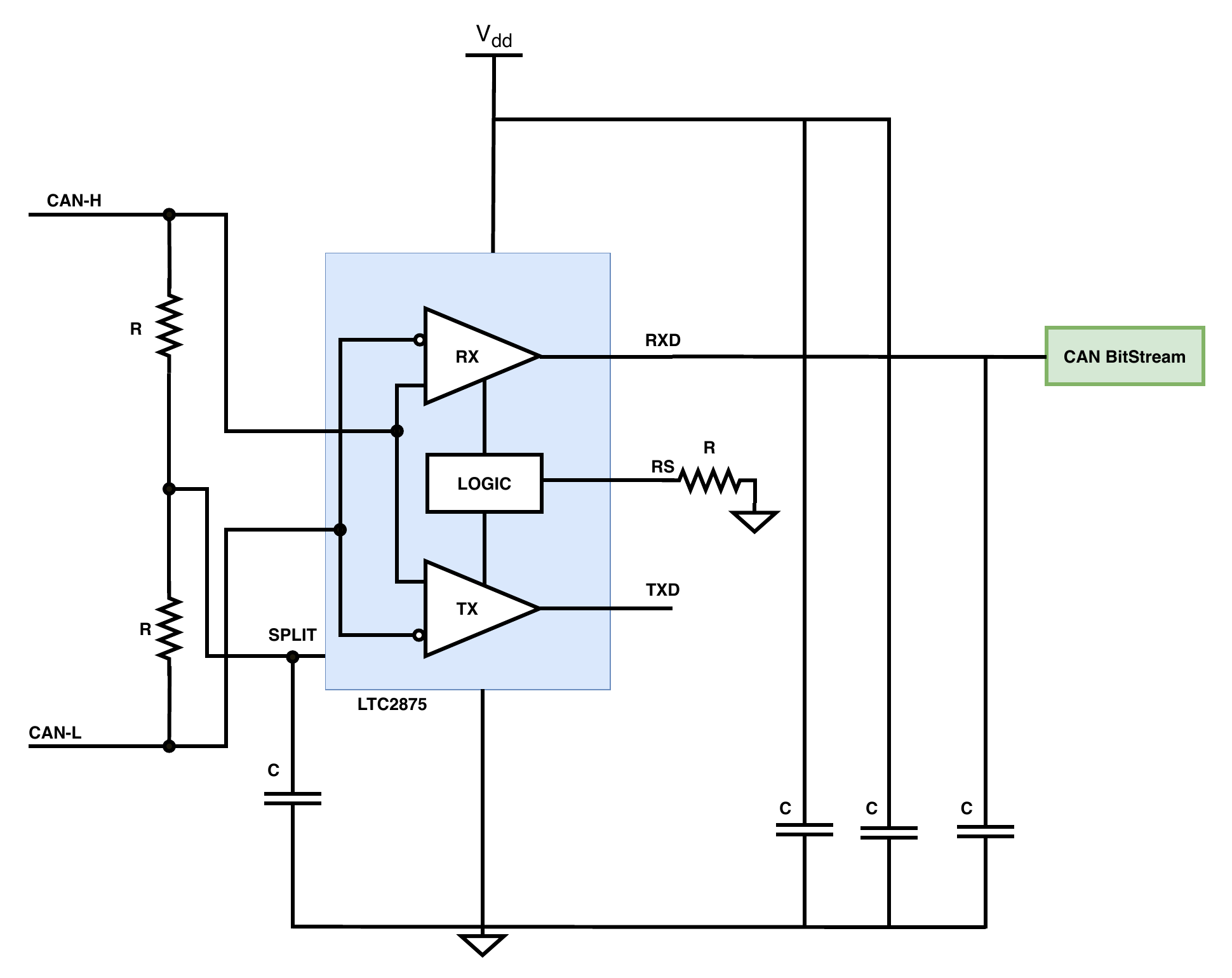}
\caption{CAN-MM Receiver - Stage 1 - SPICE Block}
\label{CANMMRXSA}
\end{figure}

Unlike the transmitter, the custom part of the \ac{CAN-MM} receiver processes data in parallel to the standard transceiver (refer to \autoref{CANMMRXSA}). The receiver includes a pass-band analog filter with a cutoff frequency set to the carrier frequency, followed by a voltage comparator with a voltage reference set to the absolute value of the noise (in this case, 100 mV). These stages form the first decode chain for \ac{CAN-MM} and are identical for both \ac{CAN} lines (see \autoref{CANMMRXSA}).

\begin{figure}[htb]
\centering
\includegraphics[width=\columnwidth]{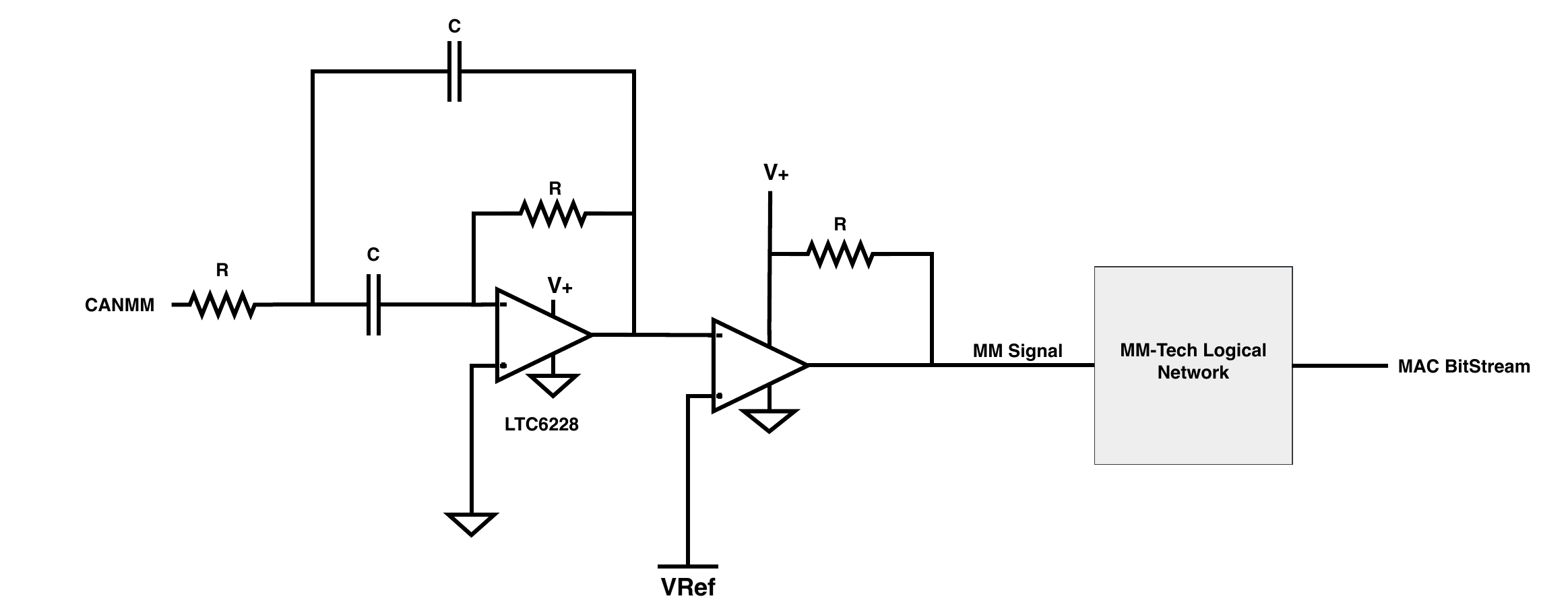}
\caption{CAN-MM Receiver - Stage 2 - SPICE Block}
\label{CANMMRXSB}
\end{figure}

In the second stage of the \ac{CAN-MM} receiver, the contribution on the two \ac{CAN} lines is collapsed together through a \texttt{NOR} port. Downstream of the \texttt{NOR} port is a custom logical network based on flip-flop counters, which is used to extract the \ac{MAC} contribution (refer to \autoref{CANMMRXSB}).

\subsection{Preliminary Hardware Implementation \label{sec:hardwareImpl}}

\begin{figure}[htb]
\centering
\includegraphics[width=\columnwidth]{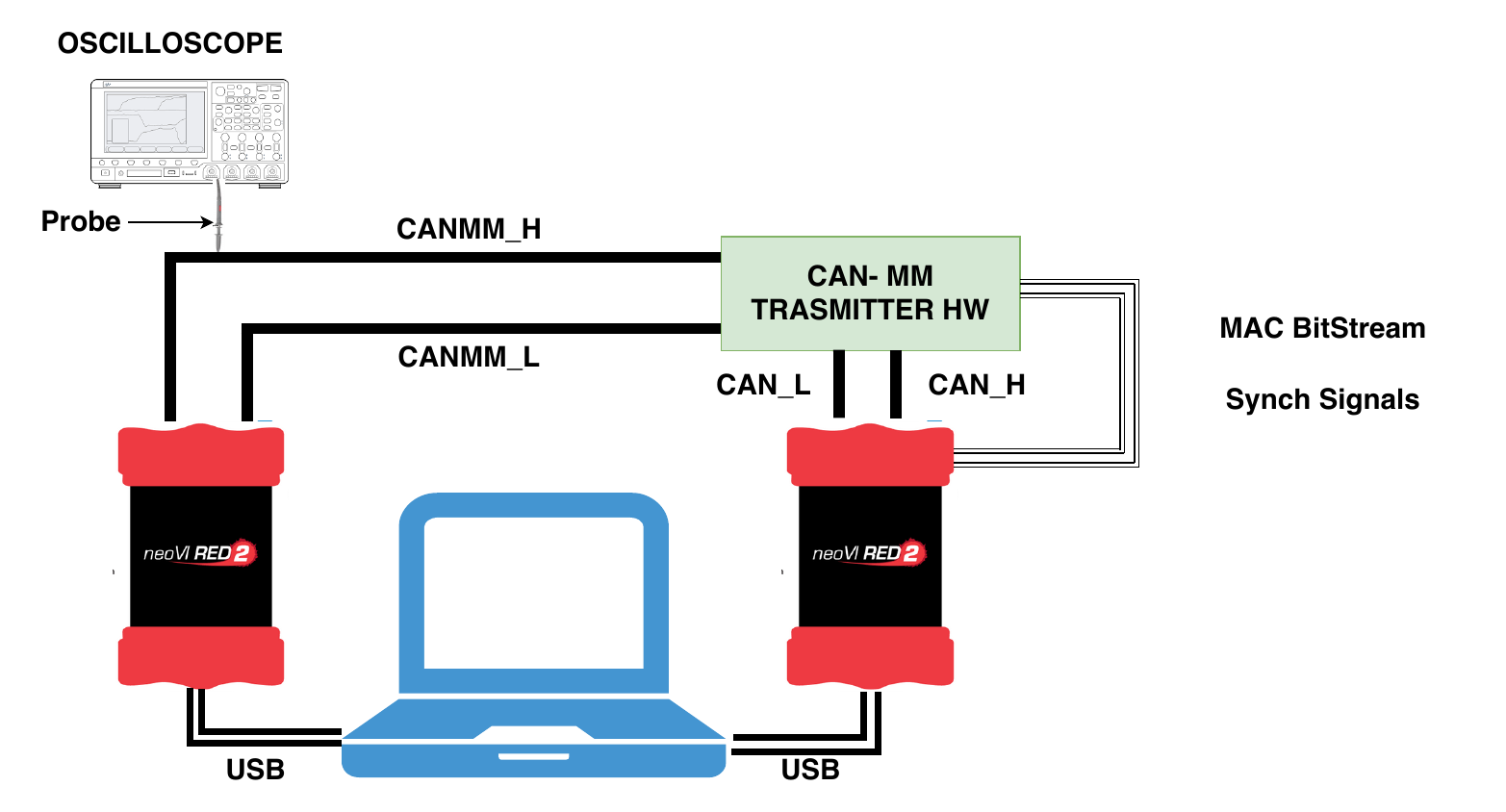}
\caption{CAN-MM Hardware Concept Scheme}
\label{fig:BENHSCH}
\end{figure}

A hardware prototype was created to enhance the validation of the \ac{CAN-MM} technology. The prototype is specifically designed to assess the functionality of the CAN-MM transmitter. It is implemented within a compact In-Loop CAN network, as illustrated in Figure \ref{fig:BENHSCH}. The primary goal of this validation is to confirm the capability of a standard receiver to receive the \ac{CAN-MM} conditioned signal accurately.

The experimental setup involves a laptop connected to a Neo VI Multi-Protocol Vehicle Interface, which oversees a custom hardware board designed for \ac{CAN-MM} operation. This board is crucial for converting the incoming \ac{CAN} signal, received through the Neo VI interface, into a \ac{CAN-MM} frame. The conversion process is directed by control signals continuously managed by the Neo VI device. Additionally, the hardware board is linked to another Neo VI device via the \ac{CAN-MM} bus, set up to function under the standard \ac{CAN} protocol. This configuration creates a closed loop with the laptop, facilitating seamless communication.

Notably, the \ac{CAN-MM} bus is deliberately designed to be open-access, enabling the intentional introduction of noise and permitting data acquisition with an oscilloscope. In the second stage of the loop-back scheme, a programmable noise source was also added to simulate the noise profile acquired during the idle operation of the engine, as previously used in the LT-Spice simulations.

\section{Experimental results}
\label{sec:results}

\begin{figure}[htb]
\centering
\includegraphics[width=0.98\columnwidth]{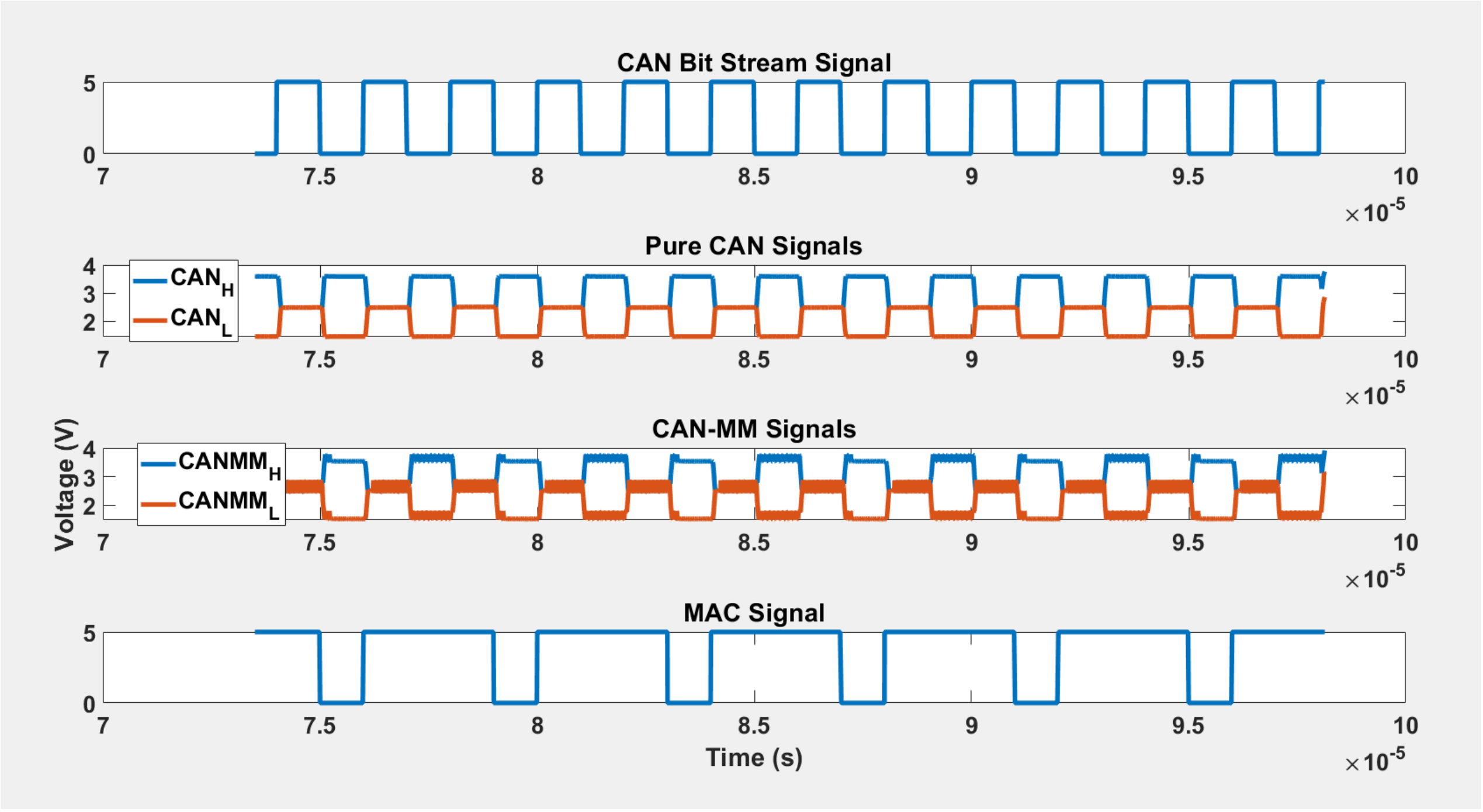}
\caption{CAN-MM transmitter output}
\label{fig:15}
\end{figure}

The collected signal diagrams, illustrated in the following figures, show the electrical signals generated by each module depicted in \autoref{fig:14}. The output signals generated by \ac{CAN-MM} node \#1 are illustrated in \autoref{fig:15}, which depicts four subplots. The blue line in the first subplot illustrates a section of a transmitted \ac{CAN} bitstream, while the second one displays the differential electrical signals. The third subplot shows the \ac{CAN-MM} electrical signals that are transmitted on \ac{CANH} and \ac{CANL}, where \ac{MAC} signal in the fourth subplot is multiplexed.

\begin{figure}[htb]
\centering
\includegraphics[width=0.98\columnwidth]{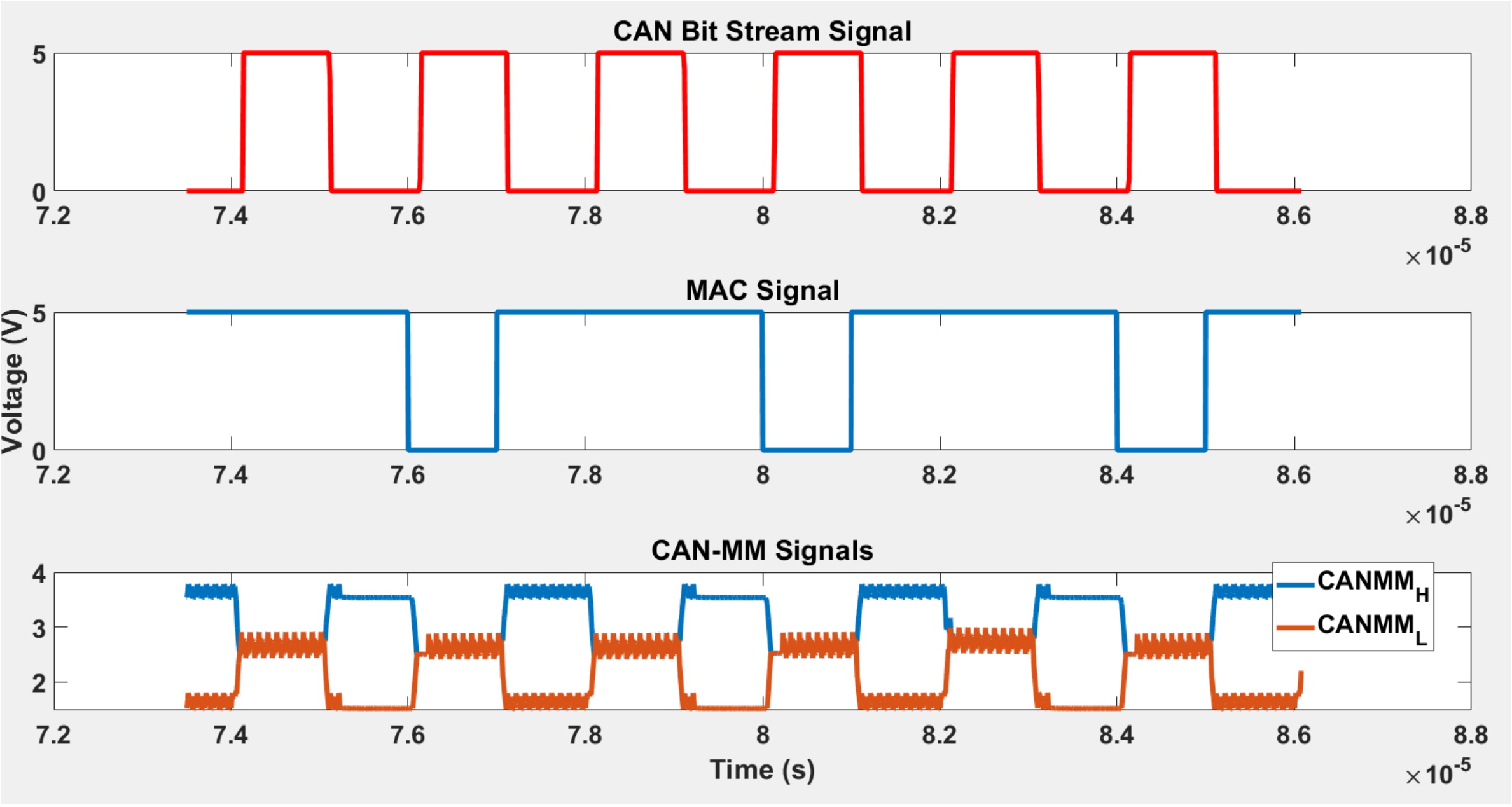}
\caption{CAN-MM receiver signals}
\label{fig:16}
\end{figure}

\autoref{fig:16} depicts the functionality of the \ac{CAN-MM} receiver in node \#2. It shows how the receiver manages the physical signal generated by the \ac{CAN-MM} transmitter and transmitted on the bus. The bottom subplot displays the received \ac{CAN-MM} physical signal through the \ac{CAN-MM} transceiver, which is identical to the signal transmitted by node \#1 in \autoref{fig:15}. The subplot in blue color is the \ac{MAC} bitstream extrapolated by the \ac{CAN-MM} decoder in node \#2, and it is the corresponding \ac{MAC} of the subplot in red color.

\begin{figure}[htb]
\centering
\includegraphics[width=0.98\columnwidth]{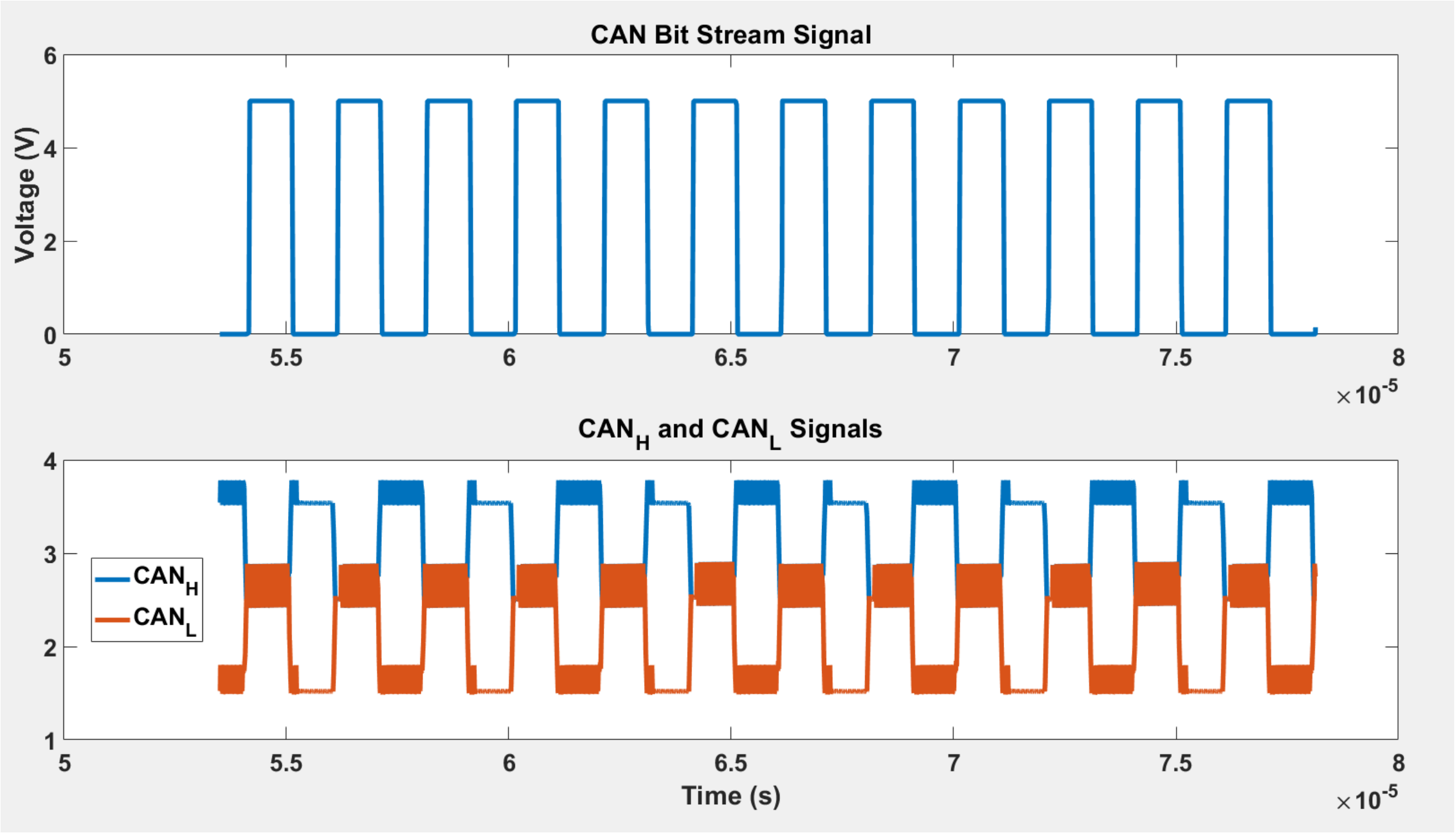}
\caption{CAN 2.0 transceiver}
\label{fig:17}
\end{figure}

To demonstrate the complete compatibility of \ac{CAN-MM} with the standard \ac{CAN} 2.0 protocol, node \#3 simulates a standard \ac{CAN} 2.0 transceiver. As shown in \autoref{fig:17}, the backward compatibility is guaranteed, as the transceiver can reconstruct the correct \ac{CAN} bitstream when it receives a \ac{CAN} frame modulated under \ac{CAN-MM} specifications. However, a standard \ac{CAN} transceiver lacks the extended hardware required to demodulate the \ac{MAC} bitstream, making it impossible to extract it.

To support a timewise analysis of the \ac{CAN-MM} to understand the potential benefits of the parallel transmission of the \ac{MAC} alongside the data payload, we computed the \ac{MAC} transmission Extra Time ($MET$), introduced by the transmission of the \ac{MAC} digest.
It depends on the \ac{MAC}'s length in bits ($MAC size$) and the selected \ac{CAN} protocol transmission time of a data bit ($\tau_{\text{dbit}}$~\cite{10075498}), as shown in equation \autoref{tb}. 
\begin{equation}\label{tb}
    MET={MAC size}*{\tau_{\text{dbit}}} 
\end{equation}

Aligning with the experimental setup in \cite{10075498}, we computed $MET$ using $\tau_{\text{dbit}}$=0.00025 ($ms$) for the \ac{CAN-FD} and $\tau_{\text{dbit}}$ equal to 0.0001 ($ms$) for the \ac{CAN-XL}.

In a \ac{CAN-FD} the $MET$ required to transmit the 64-bit \ac{MAC} digest is 16 \textmu s, as per equation \autoref{tb1}

\begin{equation}\label{tb1}
    MET={MAC size}*{\tau_{\text{dbit}}} = {64}*{0.00025}={16} (\mu s)
\end{equation}

Adopting a more traditional baud rate on \ac{CAN-FD}, 500kbps, we calculate a $\tau_{\text{dbit}}$ = 0.002($ms$). In this condition, the extra transmission time required by \ac{MAC} appended to the payload is 128 \textmu s (see \autoref{tb3}).

\begin{equation}\label{tb3}
    MET={MAC size}*{\tau_{\text{dbit}}} = {64}*{0.002}={128} (\mu s)
\end{equation}

Keeping the \ac{MAC}'s size constant, adopting the \ac{CAN-XL} protocol with a speed rate of 10Mbps, the $MET$ would be reduced to 6.4 \textmu s, which represents the best possible transmission performance by \ac{SecOC} and \ac{CANsec}, as per equation \autoref{tb2}, demonstrating that a broad adoption fo \ac{CAN-XL} would introduce faster performance.

\begin{equation}\label{tb2}
    MET={MAC size}*{\tau_{\text{dbit}}} = {64}*{0.0001}={6.4} (\mu s)
\end{equation}

Opting for \ac{CAN-MM} instead highlights a key benefit: the negligible impact on transmission times due to \ac{MAC}. This capability to maintain consistent transmission times, with or without \ac{MAC}, offers a solution to the schedulability challenges discussed by Ikumapayi et al.\cite{10075498}.
Moreover, \ac{CAN-MM} supports countermeasures on the schedulability noted by the authors of~\cite{9106836}.
The systems described in the paper adopt \ac{RMS}, a deterministic scheduling algorithm for real-time operating systems that assign priorities to tasks based on their period; the shorter the period, the higher the priority. A pivotal aspect of \ac{RMS} is its CPU utilization bound for $n$ periodic tasks, which can be calculated using the Liu \& Layland formula, \autoref{eq:tb4}, where $C_i$ is the computation time of task $i$, $T_i$ is the period of task $i$, and $U$ is the total CPU utilization. This formula ensures that if the total CPU utilization is below a certain threshold, all tasks can be scheduled to meet their deadlines, making \ac{RMS} particularly efficient for systems with hard real-time constraints.
 
 \begin{equation}\label{eq:tb4}
 U = \sum_{i=1}^{n} \frac{C_i}{T_i} \leq n(2^{\frac{1}{n}} - 1)
 \end{equation}

The transmission time of the \ac{CAN} and the \ac{MAC} might significantly contribute to $C_i$, the computational load. By reducing the transmission time, \ac{CAN-MM} directly decreases $C_i$ and, consequently, the total CPU utilization. This reduction is crucial for enhancing resilience against certain types of attacks.

To provide a general understanding, the \ac{HSM} performance metrics published by Pott~\cite{Pott2021} indicate that more than 300 clock cycles are required for \ac{MAC} verification. When considering latency, the total time is approximately 5-6 \textmu s, which parallels the time savings achieved by \ac{CAN-MM} compared to \ac{CAN-XL}. Consequently, this denotes that \ac{CAN-MM} might theoretically offer a twofold increase in the system's ability to withstand such attacks, in contrast to the conventional \ac{CAN-XL} framework where the \ac{MAC} is appended to the payload.

\begin{figure}[htb]
\centering
\includegraphics[width=0.98\columnwidth]{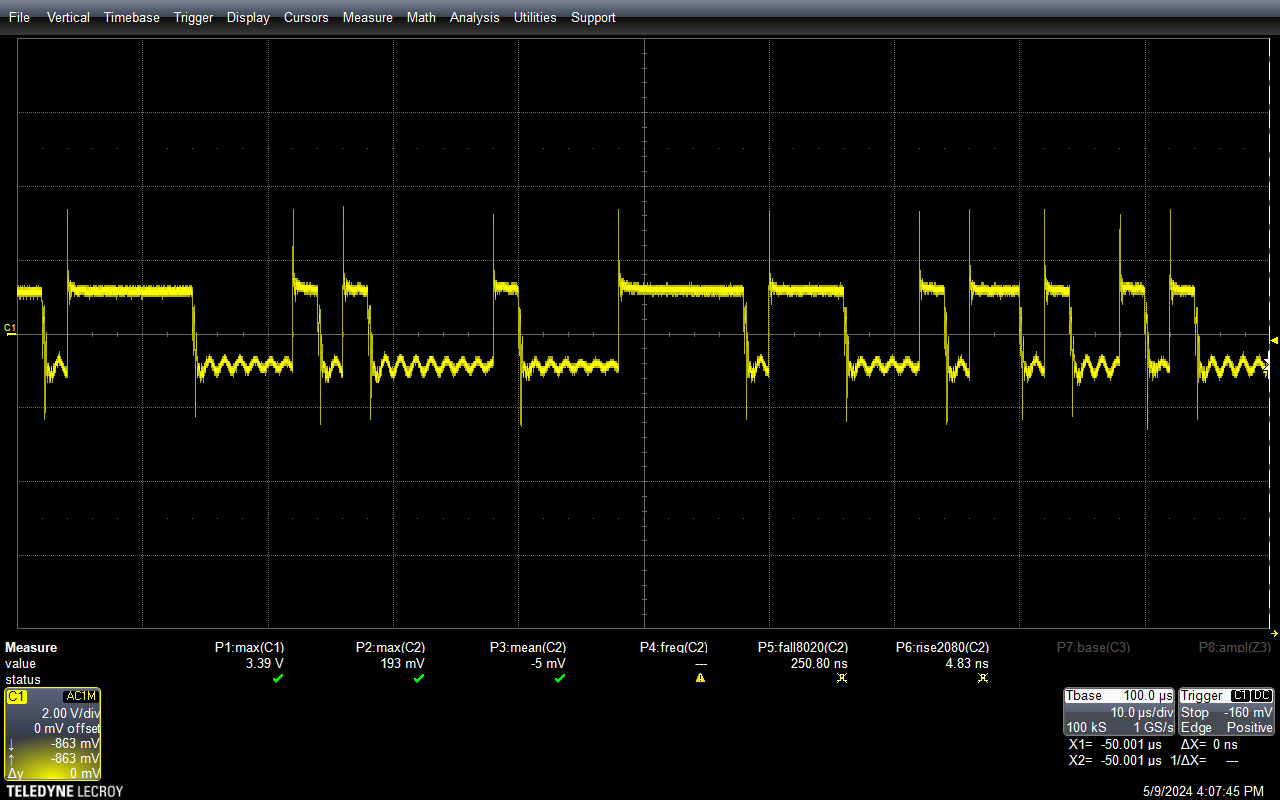}
\caption{CAN-MM-H acquired by Oscilloscope}
\label{fig:Oscil}
\end{figure}

The robustness of \ac{CAN-MM} was further validated through measures performed on the hardware implementation introduced in \autoref{sec:hardwareImpl}. These results complement the ones produced by the LT-SPICE simulations. The captured data in \autoref{fig:Oscil} portrays the real-time \ac{CAN-MM}-H bus traffic. The applied noise profile follows what has been captured from a vehicle as described in \autoref{sec:noismodel}. Within this experimental framework, the \ac{CAN-MM} transmitter effectively performs the multiplexing of the \ac{MAC} Bitstream, precisely the bit sequence \texttt{000011101011110111}, over the underlying physical \ac{CAN}-H signal. This multiplexing process is executed through the \ac{OOK} modulation technique, closely replicating the observations obtained in the simulated environment, thus confirming the robustness of the \ac{CAN-MM} system.

Moreover, the BUSMASTER \cite{BUSMASTER} tool reported error-free reception of the transmitted \ac{CAN} message. This confirms the backward compatibility of the \ac{CAN-MM} approach with conventional hardware. The multiplexed carrier of the standard transceiver is intelligently filtered out, effectively treating it as noise in the system.

\section{CAN-MM Type-B}
\label{sec:CANMMTB}

\autoref{sec:results} highlights a potential limitation in the \ac{CAN-MM} architecture when the carrier and noise frequencies align, manifesting sporadic failures in demodulating the \ac{MAC} bit-stream. While this scenario is unlikely to occur in actual situations, considering that noise amplitudes exceeding 100mV are seldom encountered, this paper introduces an advanced \ac{CAN-MM} architecture called Type-B, able to withstand scenarios where the carrier signal frequency matches the noise. \ac{CAN-MM} Type-B ensures additional robustness to noise across all frequency bands without risking data corruption.

\begin{figure}[htb]
    \centering
    \includegraphics[width=0.98\columnwidth]{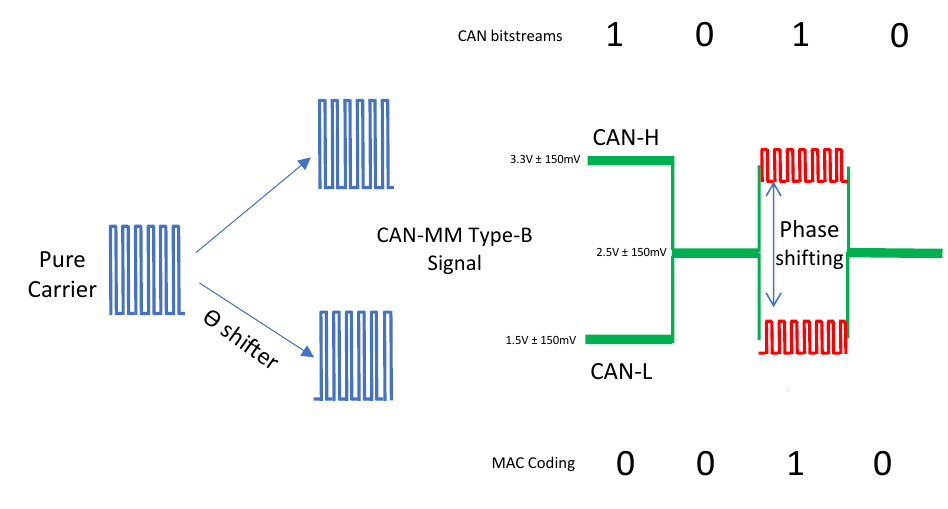}
    \caption{CAN-MM Type-B physical signals scheme}
    \label{fig:10}
 \end{figure}

\begin{figure}[hb]
    \centering
   \includegraphics[width=0.98 \columnwidth]{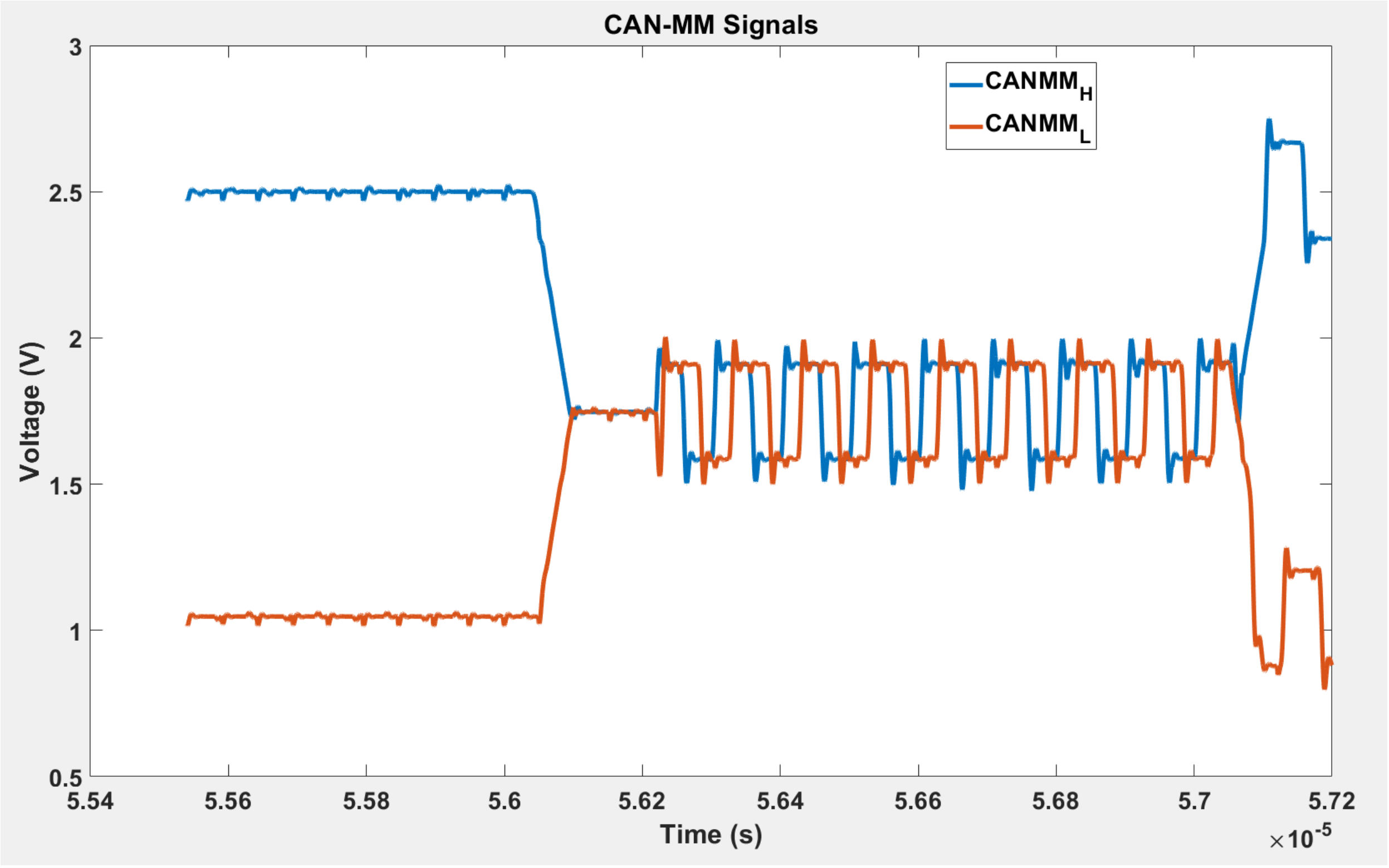}
    \caption{CAN-MM Type-B Physical Signal with the shifted carrier on CAN-L}
   \label{fig:CANMMTBS}
 \end{figure}

The \ac{CAN-MM} Type-B physical signals scheme incorporates \ac{CPSM}~\cite{8591776} as depicted in \autoref{fig:10}. The \ac{CPSM} carrier varies between \ac{CANH} and \ac{CANL}, causing a phase shift ranging from 90° to 270°. The proposed design sets the phase modulation to 90° for \ac{CANL} as depicted in \autoref{fig:CANMMTBS}.  

    \begin{figure*}[htb]
    \centering
    \includegraphics[width=0.98\textwidth]{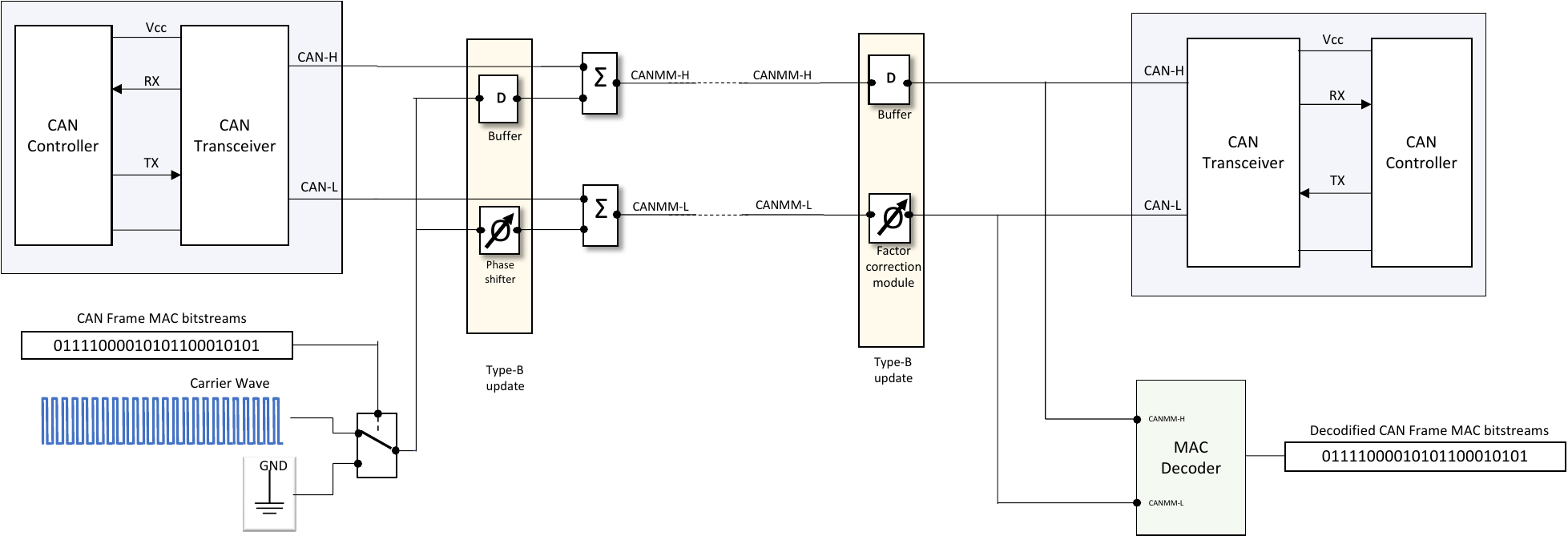}
   \caption{CAN-MM Type-B Transmitter\& Receiver Block scheme}
   \label{fig:11}
 \end{figure*}

\begin{figure}[htb]
    \centering
   \includegraphics[width=0.98\columnwidth]{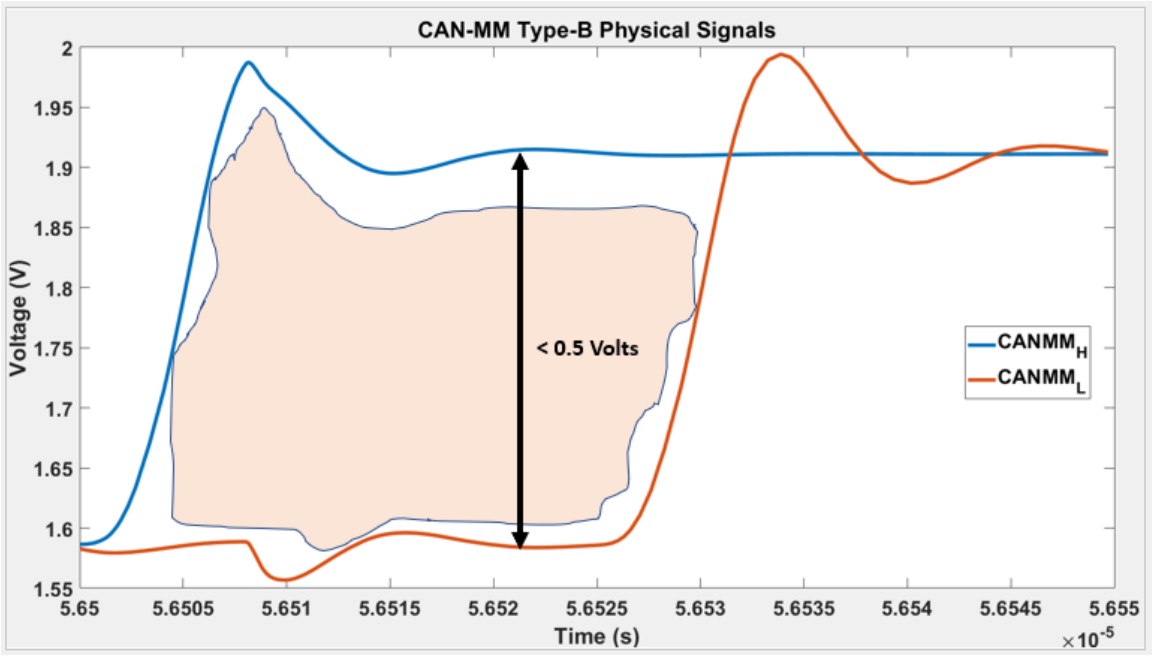}
    \caption{Critical Area due to shifting phase for codification correctness}
   \label{fig:CANMMTBFD}
\end{figure}

The additional phase-shifting can result in incorrect codification, particularly if the differential voltage in the red area depicted in \autoref{fig:CANMMTBFD} exceeds the 0.5V threshold. To overcome this limitation, an additional re-phaser stage represented by the orange area in the receiver reported in \autoref{fig:11} reverses the \ac{CPSM} applied by the \ac{CAN-MM} Type-B transmitter. This block is placed at the very beginning of the reception process. Once the re-phasing is completed, the standard \ac{CAN-MM} receiver, which includes the standard \ac{CAN} transceiver and the \ac{CAN-MM} decoder, work in parallel to extract their respective data from the re-phased frame. 
 
The additional protection to noise of \ac{CAN-MM} Type-B across all frequency ranges comes with the cost of adding an upstream hardware re-phaser block to the \ac{CAN} transceiver when it functions as a receiver. 

\begin{figure}[hb]
    \centering
   \includegraphics[width=0.98 \columnwidth]{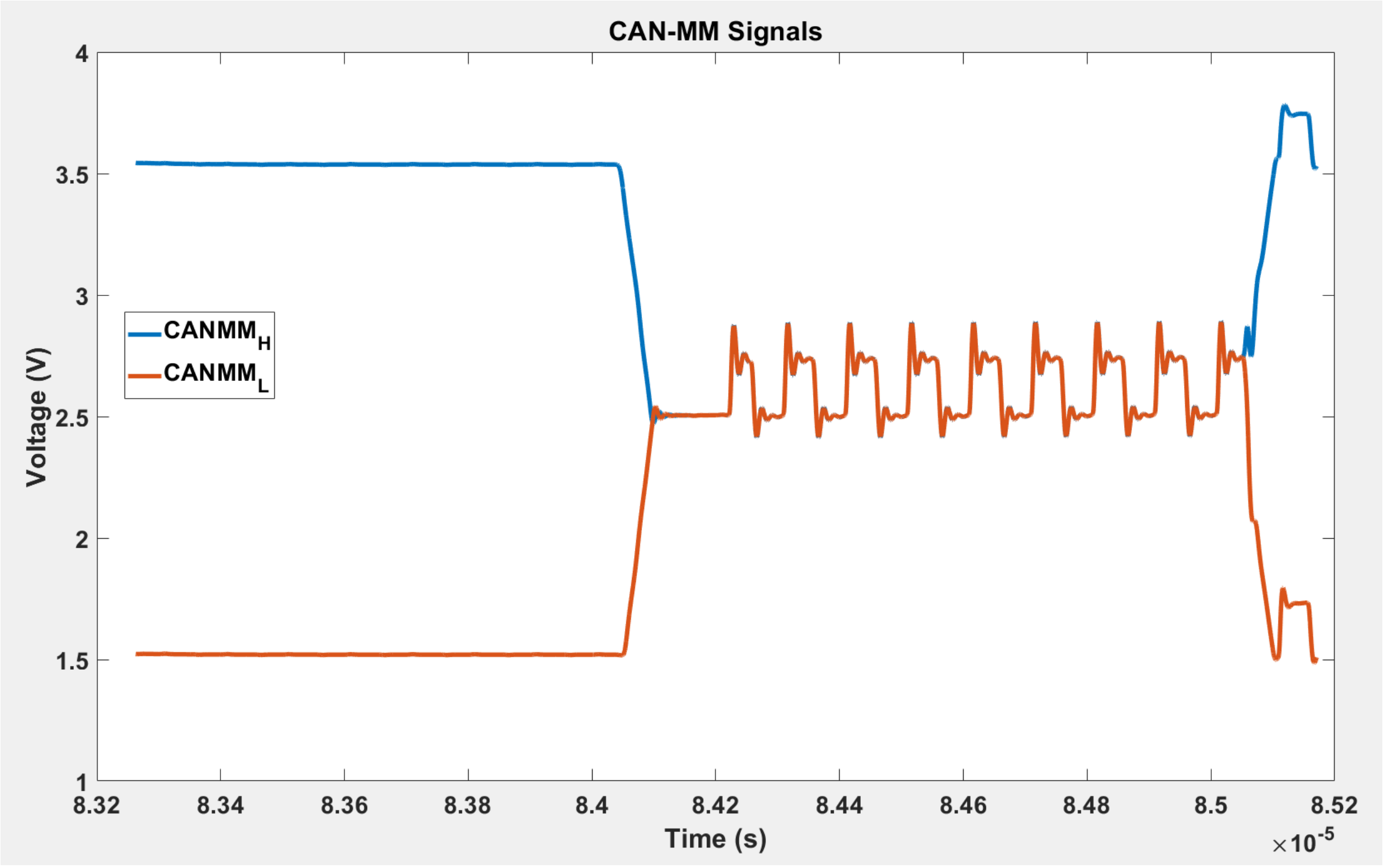}
    \caption{CAN-MM Type-B filter scheme}
   \label{fig:CANMMTBRS}
 \end{figure}

 An LT-Spice model was developed to validate the robustness of the \ac{CAN-MM} Type-B architecture (\autoref{fig:CANMMTBS}).  
 \ac{MAC} code 1 is encoded by adding a carrier with a shifting phase on \ac{CANL}, allowing for greater robustness during decoding activities. However, in certain regions, the phase shifting can cause the differential voltage between these signals to exceed the 0.5V limit. Thus, as shown in \autoref{fig:CANMMTBRS}, the signal is shifted back before decoding, obtaining full synchronization. 
 
\begin{figure}[htb]
    \centering
   \includegraphics[width=0.98 \columnwidth]{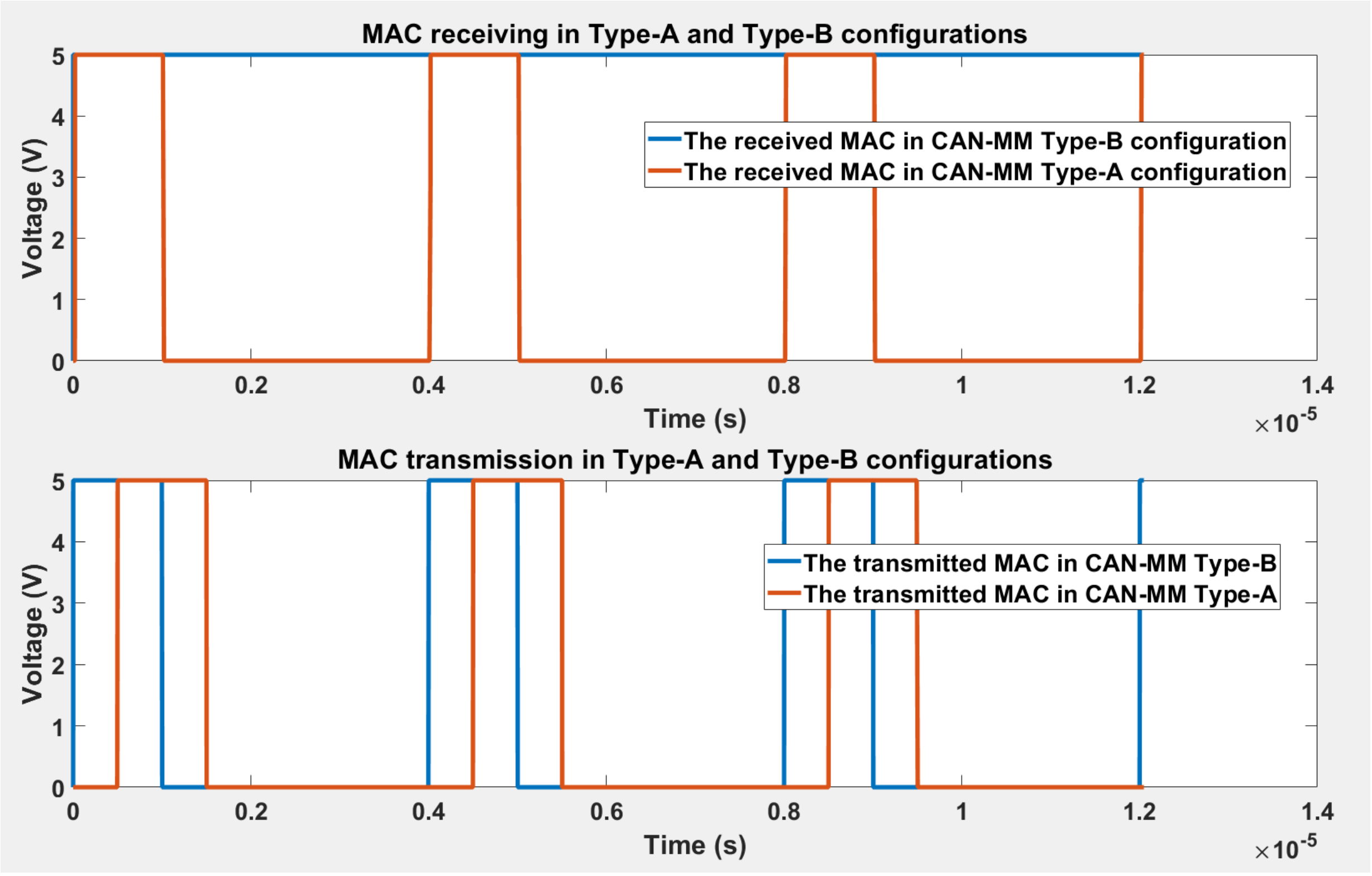}
    \caption{CAN-MM Type-A vs. CAN-MM Type-B Noise capability performances}
   \label{fig:NANB}
 \end{figure}

 \autoref{fig:NANB} presents a comparative comparison between \ac{CAN-MM} and \ac{CAN-MM} Type-B to validate the design robustness. This experiment applies a noise signal with a 140mV amplitude to the original \ac{CAN-MM} architecture model. The investigation is completed only for completeness since the resulting signal is clearly out of specification. As a result of the high noise level, the receiver could not extract the correct \ac{MAC} bit-stream, and the output was a \ac{MAC} bit-stream stuck to 1. However, in the case of \ac{CAN-MM} Type-B, despite a noise signal with an amplitude of 200mV, the receiver correctly decoded the \ac{MAC} stream.

\begin{figure}[htb]
\centering
\includegraphics[width=0.98\columnwidth]{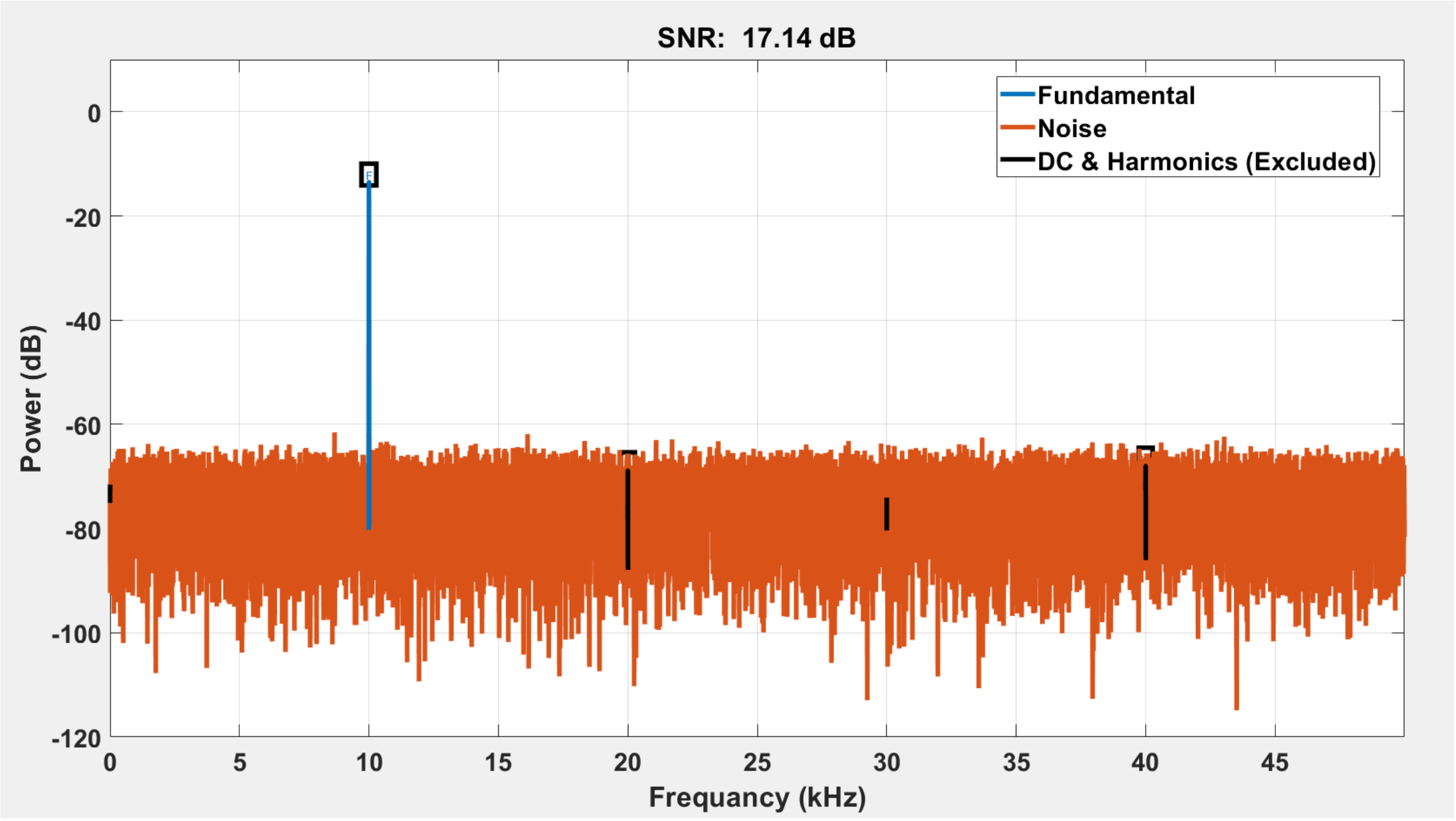}
\caption{SNR CAN-MM TypeB Graph}
\label{fig:SNRB}
\end{figure}

By referring to \autoref{fig:SNRB}, we have calculated the signal-to-noise ratio (\ac{SNR}) for this scenario to be approximately 17.32 dB. This high \ac{SNR} value underscores the signal's robustness, affirming its clear distinction from the surrounding background noise.

\section{Security Analysis}
\label{sec:SecAnl}

This section delves into the security aspects of the \ac{CAN-MM} architecture, particularly addressing attack models outlined in \autoref{sec:CAN-overview}. 

The main objective of \ac{CAN-MM} is to support a full \ac{CAN} 2.0 vehicle network security by embedding a \ac{SecOC} compatible \ac{MAC} code within each payload frame, matching the same level of protection of \ac{CAN-FD}. 
Moreover, it supports security against threats such as \ac{MitM} and replay attacks due to the presence of the \ac{MAC} mechanism that neutralizes those types of attacks. This capability also includes the more recent Janus attack, as described by the author~\cite{Tindell:2024aa}.

\ac{CAN-MM} may also neutralize Cloak attacks by maintaining payload integrity, even amidst bit modifications. Leveraging the sample rate of two receivers will be more complex if the attacker also must coherently switch the modulated \ac{MAC}. Such complexity will narrow the timing window where the attack is effective, as discussed in the original paper~\cite{Yue:2021aa}.

When a significant challenge arises when the system is overwhelmed by an excessive number of \ac{MAC}s that need to be validated~\cite{9106836}, the validation process demands intensive cryptographic computations, potentially compromising the system's ability to adhere to real-time deadlines. This issue becomes particularly acute with the influx of numerous fraudulent \ac{MAC}s. The \ac{CAN-MM} system introduces enhanced security measures against those kinds of attacks. 

\section{Conclusion}
\label{sec:conclusions}

This paper presented an efficient solution to mitigate security concerns within the automotive domain's fundamental communication protocol, the \ac{CAN}. The proposed solution, \ac{CAN-MM}, facilitates the transmission of \ac{MAC} payloads in standard \ac{CAN} to complement any security schemas based on it efficiently. The support of the \ac{MAC} transmission also safeguards the automotive communication system against \ac{MitM} and replay attacks.

The \ac{CAN-MM} architecture, developed to upgrade communication hardware for upcoming security regulations, maintains compatibility with existing \ac{CAN} devices, avoiding the necessity for a complete system or vehicle architecture overhaul. This hybrid networking capability offers flexibility to designers, minimizing the requirement for updating electronic components to the new generation and thereby reducing the cost of transitioning a vehicle fleet into the cyber-secure domain. 

Additionally, an improved Type-B version of \ac{CAN-MM} addresses potential demodulation issues without sacrificing backward compatibility. While this modified version may compromise some degree of backward compatibility, the applied modulation technology to the \ac{CAN} protocol can be extended not only to version 2.0 but also to other existing versions that already incorporate the \ac{MAC}.

\ifCLASSOPTIONcaptionsoff
  \newpage
\fi

\bibliographystyle{IEEEtran}
\bibliography{007_X+bibliography+CANMM}

\begin{IEEEbiography}
[{\includegraphics[width=1in,height=1.25in,clip,keepaspectratio]{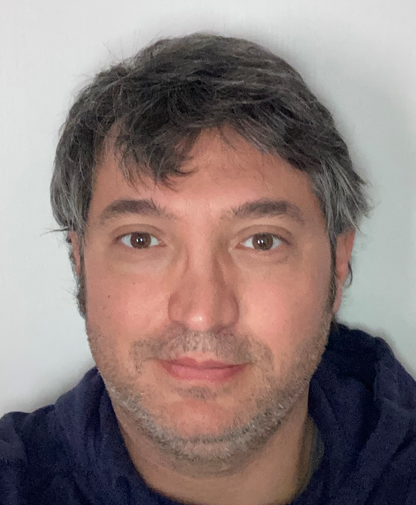}}]
{Franco Oberti}
Franco Oberti (student, IEEE '20) received an M.Sc. degree in computer engineering from the Politecnico di Torino, Torino, Italy, in 2007. He started working in Dumarey Softronix (former General Motors Powertrain Europe) in 2007, where he held different positions. In 2016, he received a Master's certificate in Advanced Cybersecurity from Standford University. Currently, he is part of the Product Security Office in Dumarey Softronix. By 2021, he was also an Industry Ph.D. student candidate. His current research interests include cybersecurity applied to embedded road vehicle systems.
\end{IEEEbiography}

\begin{IEEEbiography}[{\includegraphics[width=1in,height=1.25in,clip,keepaspectratio]{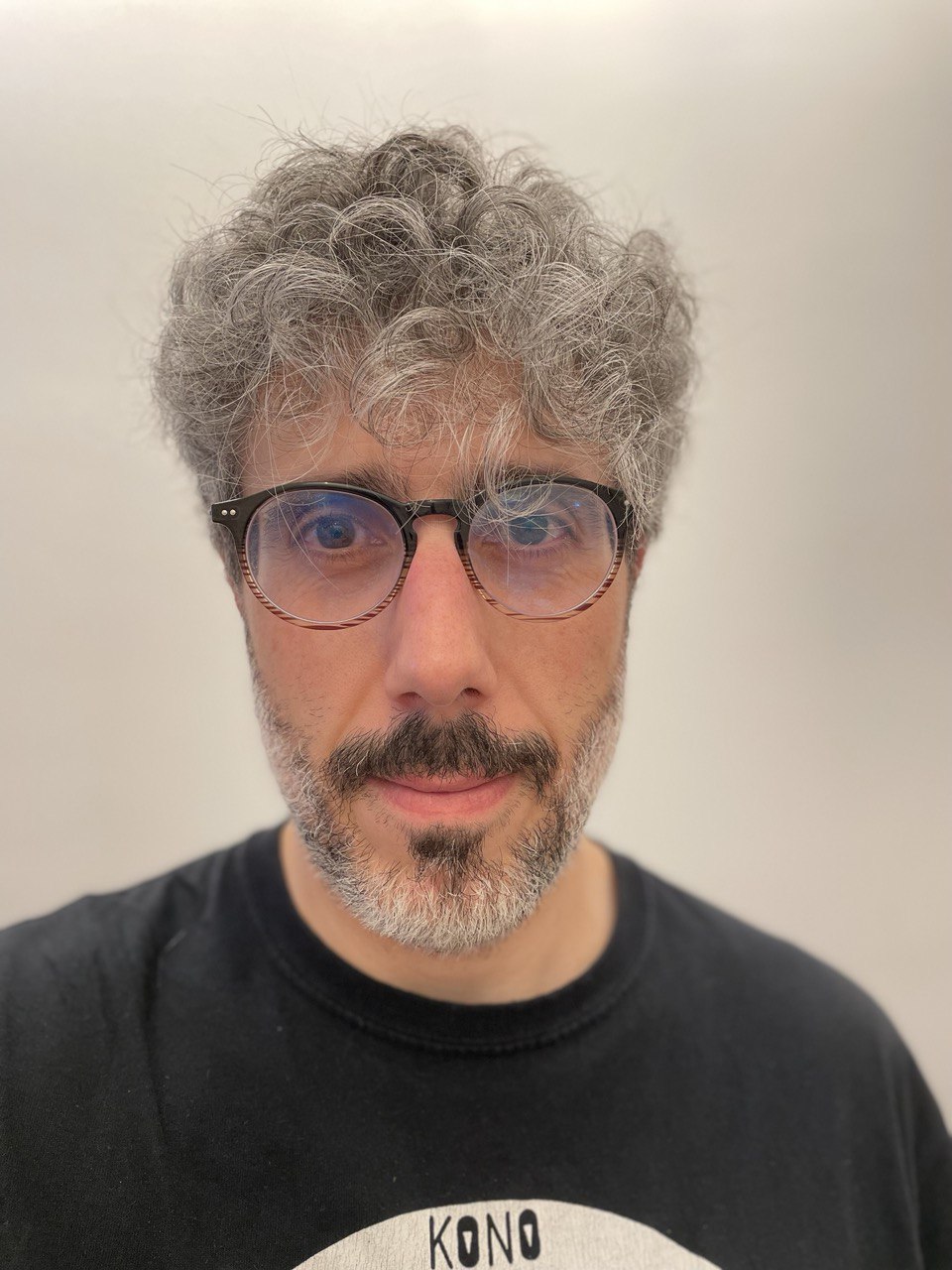}}]{Alessandro Savino}(M'14, SM'22) is an Associate Professor at the Department of Control and Computer Engineering at Politecnico di Torino (Italy). He holds a Ph.D. (2009) and an M.S. equivalent (2005) in Computer Engineering and Information Technology from the Politecnico di Torino in Italy. Dr. Savino's research contributions include Approximate Computing, System Reliability, Neuromorphic Computing, Safety-Critical Systems, Software-Based Self-Test, and Bioinformatics. He has been part of the program and organizing committee of several IEEE and INSTICC conferences and has served as a reviewer of IEEE conferences and journals. \end{IEEEbiography}

\begin{IEEEbiography}[{\includegraphics[width=1in,height=1.25in,clip,keepaspectratio]{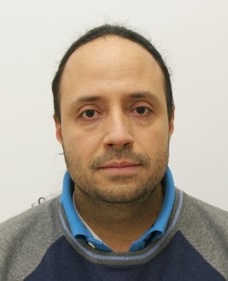}}]{Ernesto Sanchez}
He received a degree in electronic engineering from Universidad Javeriana, Bogota, Colombia, in 2000 and a Ph.D. in computer engineering from the Politecnico di Torino, Italy, in 2006, where he is currently an Associate Professor in the Department of Control and Computer Engineering. His research interests include microprocessor testing, hardware security, and DNN reliability.
\end{IEEEbiography}

\begin{IEEEbiography}[{\includegraphics[width=1in,height=1.25in,clip,keepaspectratio]{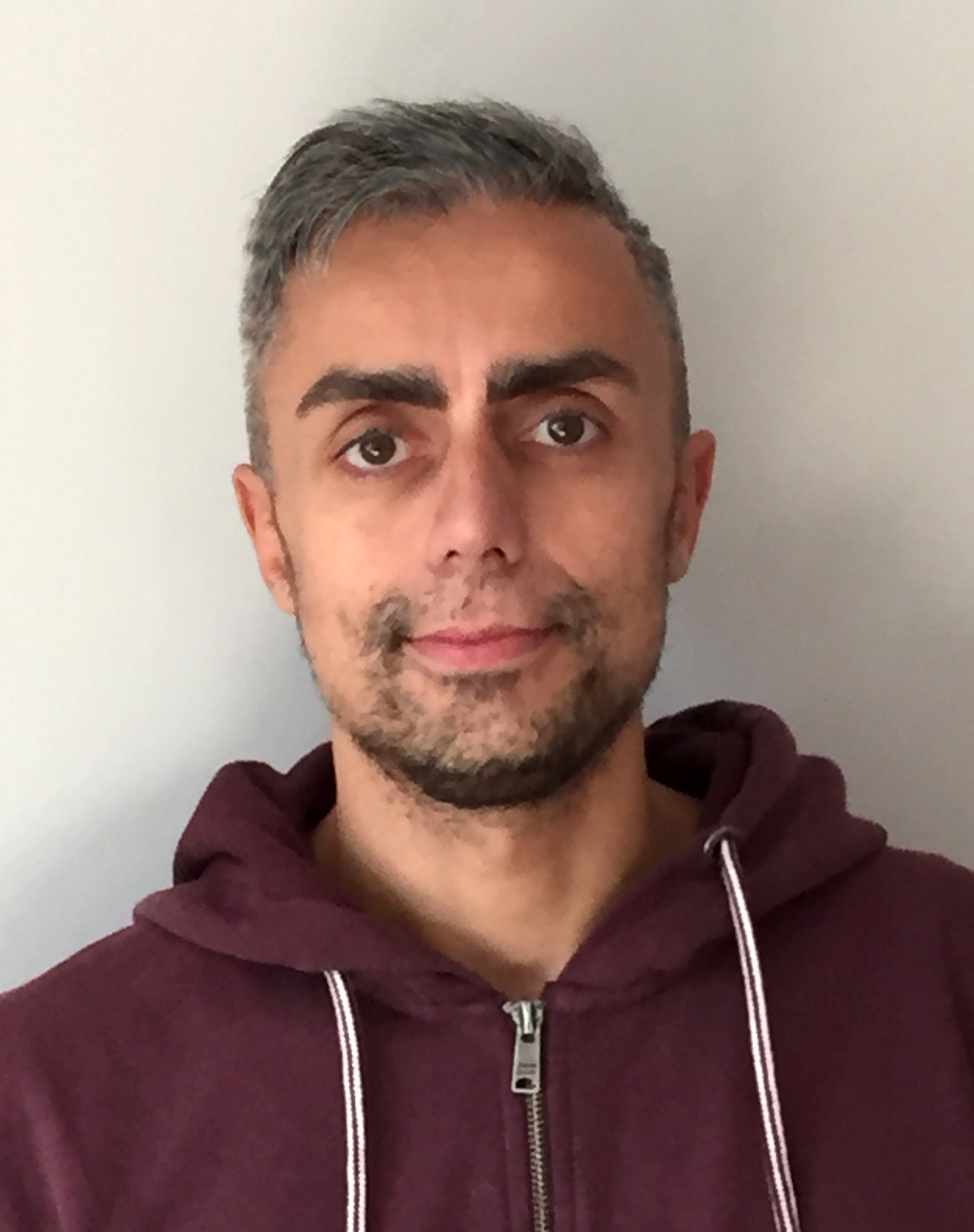}}]{Stefano Di Carlo}
(SM'00-M'03-SM'11) received an M.Sc. degree in computer engineering and a Ph.D. in information technologies from Politecnico di Torino, Italy, where he is a full professor. His research interests include computer architecture reliability, safety, and security. He has coordinated several national and European research projects. Di Carlo has published over 250 papers in peer-reviewed IEEE and ACM journals and conferences. He regularly serves on the Organizing and Program Committees of major IEEE and ACM conferences. He is a Golden Core member of the IEEE Computer Society and a senior member of the IEEE.
\end{IEEEbiography}

\begin{IEEEbiography}
[{\includegraphics[width=1in,height=1.25in,clip,keepaspectratio]{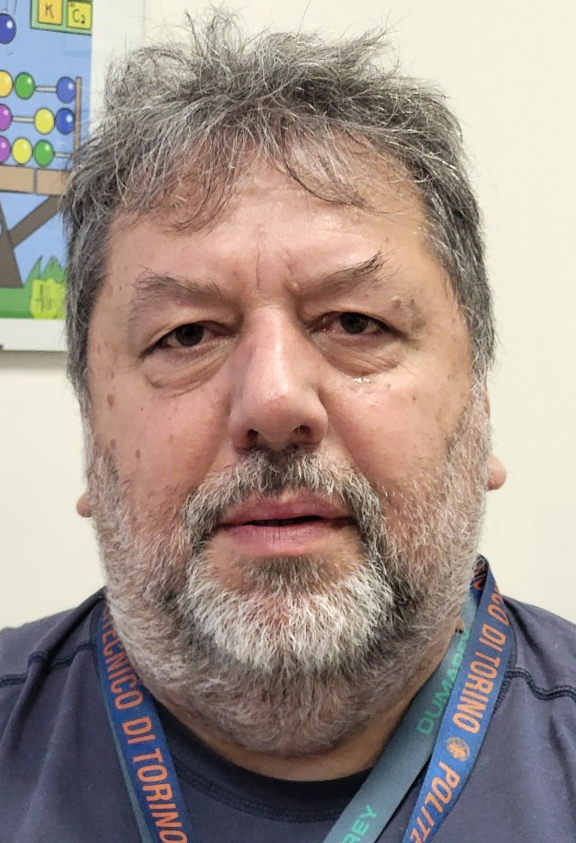}}]
{Filippo Parisi}
Filippo Parisi, aka albix, received a degree in electronic engineering from Politecnico of Turin, Turin, Italy, in 1992. As head of electronics at Dumarey Softronix, he leads the development of electronics, firmware, and virtualization for testing applied to hard real-time, safety-critical automotive embedded control systems. He held several positions in multinational automotive companies such as FIAT Research Center, FIAT-GM-Powertrain JV, and General Motors for more than 30 years. 

\end{IEEEbiography}
\begin{IEEEbiography}[{\includegraphics[width=1in,height=1.25in,clip,keepaspectratio]{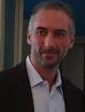}}]{Paolo Casasso}
Paolo Casasso received a degree in electrical engineering from Politecnico of Turin, Turin, Italy, in 2000. As a manager in Dumarey Softronix, he leads the hardware development team specializing in automotive embedded control systems. He held several positions in multinational automotive companies such as FIAT-GM-Powertrain JV and General Motors for more than 20 years.”
\end{IEEEbiography}

\vfill

\end{document}